\documentclass[prc,aps,amsfonts,dvips,twocolumn,floatfix]{revtex4}
\usepackage{graphics,epsfig,amssymb,color}
\usepackage{natbib}
\usepackage{ulem}
\definecolor{orange}{cmyk}{0,0.61,0.87,0}

\begin{document}

\title{
Low-lying spectroscopy of a few even-even silicon isotopes investigated by means of the multiparticle-multihole
Gogny energy density functional}
\author{N. Pillet$^{a}$, V.G. Zelevinsky$^{b}$, M. Dupuis$^{a}$, J.-F. Berger$^{a}$,
 and J.M. Daugas$^{a}$}

\bigskip

\affiliation{
$^{(a)}$ CEA, DAM, DIF, F-91297 Arpajon, France \\
$^{(b)}$ Department of Physics and Astronomy and National Superconducting Cyclotron Laboratory,
Michigan State University, East Lansing, Michigan 48824, USA }

\date{\today}

\def\fid{\vert\phi >}
\def\fig{< \phi\vert}
\def\psid{\vert\Psi>}
\def\psig{<\Psi\vert}
\def\psid{\vert\Psi>}
\def\psig{<\Psi\vert}
\def\dspt{\displaystyle}

\begin{abstract}
A multiconfiguration microscopic method has been applied with the Gogny effective interaction to the calculation
of low-lying positive-parity states in even-even $^{26-32}$Si isotopes. The aim of the study is to compare the
results of this approach with those of a standard method of GCM type and to get insight into the predictive power
of multiconfiguration methods employed with effective nucleon-nucleon force taylored to mean-field calculations.
It is found that the multiconfiguration approach leads to an excellent description of the low-lying spectroscopy of
$^{26}$Si, $^{28}$Si and $^{32}$Si, but gives a systematic energy shift in $^{30}$Si. A careful analysis of this
phenomenon shows that this discrepancy originates from too large proton-neutron matrix elements
supplied by the Gogny interaction at the level
of the approximate resolution of the multiparticle-multihole configuration mixing method done in the present study. These
proton-neutron matrix elements enter in the definition of both single-particle orbital energies and coupling matrix
elements. Finally, a statistical analysis of highly excited configurations in
$^{28}$Si is performed, revealing exponential convergence in agreement with previous work in the context of the
shell model approach. This latter result provides strong arguments towards an implicit treatment of highly excited
configurations.

\end{abstract}


\maketitle

\section{Introduction}\label{intro}

Mean field approaches provide a reliable foundation for an approximate solution of the nuclear many-body problem.
Nowadays, a lot of efforts is applied to move beyond the mean field approximation and account for missing correlations.
Special attention is paid to the restoration of symmetries broken in mean field approaches \cite{ring,anguiano1,anguiano2,bender,bender2,egido,egido2,egido3,delaroche},
for example using projection techniques. An alternative way is to develop a theory in which the trial wave functions
preserve certain symmetries. In particular, this can be achieved by multiconfiguration methods widely used in various
applications, including atomic, molecular, and condensed matter physics. When the interaction is known, this kind
of approach provides a very accurate description of a system. In a previous work \cite{pillet}, we
proposed in the nuclear physics context a variational approach based on multiparticle-multihole ($mp-mh$) proton and neutron
configuration mixings that uses the two-body finite-range density-dependent Gogny interaction in a fully microscopic way.
As a first step, the method was applied in a particular case:
pairing correlations (proton-proton and neutron-neutron) were investigated for the ground states of several even-even
tin isotopes. A pioneering work using the Skyrme SIII interaction for the mean field part and a schematic contact
interaction for the residual part has been directed to the description of $K$-isomers in the $^{178}$Hf mass region \cite{pillet1}.

The main objective of the present study is twofold: first, to test the ability of $mp-mh$ multiconfiguration approach to describe
low-lying nuclear states and second, to discuss the statistical features of highly excited configurations in nuclei. With this
aim in view, a few silicon isotopes have been chosen and their excited states calculated with the $mp-mh$ approach. Concerning
highly excited configurations, the leading idea is to investigate whether exponential convergence behavior, revealed in
the standard shell model (SM) approach \cite{ec1,ec2,ec3,ec4,ec5}, also emerges from variational ($mp-mh$) configuration
mixing methods \cite{pillet} in which single particle excitations are not restricted to a single major shell.

As practical calculations inevitably require some truncation of the orbital space and order of excitation, the $mp-mh$ method
proposes a promising scheme that allows one to predict the energies of low-lying states in a very accurate way. The possible use of
statistical properties of highly excited states, which display generic signatures of quantum chaos close to random matrix
theory, drastically reduces the sizes of the explicit diagonalizations. In the literature, one finds other
approaches proposing different algorithms as, for example, the density matrix renormalization group \cite{dmrg1,dmrg2,dmrg3,dmrg4,dmrg5}
or Monte Carlo techniques \cite{carlo}, selecting the most relevant configurations
for the description of many-body states.

The paper is organized as follows. The main characteristics of the $mp-mh$ multiconfiguration approach applied in this
study are presented in Section \ref{section1}. In Section \ref{section2}, the results obtained with the $mp-mh$ approach
for low-lying states in $^{26-32}$Si are presented and compared with those derived from a five-dimensional approximate
Generator Coordinate Method (GCM) approach (subsections \ref{section2A} and \ref{lls}).
Differences between theoretical and experimental results are discussed. In particular, the systematic energy shift found
in $^{30}$Si is analyzed in terms of proton-neutron matrix elements. Subsection \ref{pn} highlights the crucial
role played by the residual interaction between protons and neutrons.
Section \ref{section3} is devoted to the analysis of the statistical properties of highly excited configurations, taking $^{28}$Si
as an example. Conclusions and perspectives are given in Section \ref{section4}.

\section{Multiparticle-multihole configuration mixing approach}\label{section1}

In this part, we discuss a few characteristics of the $mp-mh$ configuration mixing approach. A general derivation
of the corresponding formalism in nuclear physics, with two-body density-dependent interactions, has been introduced
in Ref. \cite{pillet}. It is worth to recall here the basics of
the method not only to fix notations but also to provide an alternative formulation of equations in terms of a
``core + valence space" description.

In the $mp-mh$ configuration mixing method, the effective Hamiltonian is defined as a functional
\begin{equation}
\hat{H}(\rho)= \hat{K} + \hat{V}(\rho)
\label{eq1}
\end{equation}
of the single-particle density matrix $\rho$. In Eq. (\ref{eq1}), the Hamiltonian contains a kinetic term $\hat{K}$
(which includes the one-body center-of-mass corrections) and a two-body density-dependent potential term
$\hat{V}(\rho)$ (which includes the Coulomb potential for protons as well as the two-body center-of-mass corrections).
In our study, the D1S Gogny interaction \cite{gogny} is adopted.

The trial wave functions $\vert \Psi \rangle$ that describe nuclear stationary states are expressed as linear
combinations
\begin{equation}
\vert \Psi \rangle = \sum_{\alpha_{\pi} \alpha_{\nu}} A_{\alpha_{\pi} \alpha_{\nu}}
\vert \phi_{\alpha_{\pi}} \phi_{\alpha_{\nu}} \rangle
\label{eq2}
\end{equation}
of direct products
\begin{equation}
\vert  \phi_{\alpha_{\pi}} \phi_{\alpha_{\nu}} \rangle = \vert \phi_{\alpha_{\pi}} \rangle \otimes
\vert \phi_{\alpha_{\nu}} \rangle
\label{eq3}
\end{equation}
of proton and neutron Slater determinants, $ \vert \phi_{\alpha_{\pi}}\rangle$ and
$\vert \phi_{\alpha_{\nu}} \rangle$ respectively, containing {\it{a priori}} any multiple {\sl p-h} excitations that respect
conserved quantum numbers.

Eq. (\ref{eq2}) contains two sets of unknown parameters, the mixing coefficients $A_{\alpha_{\pi} \alpha_{\nu}}$
and the single-particle orbitals. Both are determined by applying a variational principle to a functional
$\cal{F}(\rho)$ related to the total energy of the system,
\begin{equation}
{\cal{F}}(\rho)= \langle \Psi \vert \hat{H}(\rho) \vert \Psi \rangle - \lambda \langle \Psi \vert \Psi \rangle -
\sum_{i} \lambda_{i} Q_{i},
\label{eq3b}
\end{equation}
where $\lambda$ and $\lambda_{i}$ are Lagrange multipliers and $Q_{i}$ possible additional constraints that we will leave out in the following.
One assumes that the one-body density $\rho$ entering the effective Hamiltonian $\hat{H}(\rho)$ is the correlated one:
$\rho=\langle \Psi \vert \hat{\rho} \vert \Psi \rangle$.

The minimization of ${\cal{F}}(\rho)$ with respect to the $A_{\alpha_{\pi} \alpha_{\nu}}$ leads
to a non-linear secular equation that is equivalent to a diagonalization problem in the multiconfigurational space
of a Hamiltonian matrix ${\cal{H}}$,
\begin{equation}
\sum_{\alpha'_{\pi} \alpha'_{\nu}} {\cal{H}}_{\alpha_{\pi} \alpha_{\nu},\alpha'_{\pi} \alpha'_{\nu}}
A_{\alpha'_{\pi} \alpha'_{\nu}}= \lambda A_{\alpha_{\pi} \alpha_{\nu}}.
\label{eq4}
\end{equation}
In Eq. (\ref{eq4}), the matrix $ {\cal{H}}$ contains contributions of the Hamiltonian $ \hat{H}(\rho) $ and of rearrangement terms
that come from the density dependence of the interaction,
\begin{equation}
{\cal{H}}_{\alpha_{\pi} \alpha_{\nu},\alpha'_{\pi} \alpha'_{\nu}}
= \langle \phi_{\alpha_{\pi}} \phi_{\alpha_{\nu}}  \vert \hat{H}(\rho) + \sum_{mn \tau}
{\cal{R}}^{\tau}_{mn} a^{+}_{\tau m} a_{\tau n} \vert \phi_{\alpha'_{\pi}} \phi_{\alpha'_{\nu}} \rangle,
\label{eq5}
\end{equation}
where the summation over $\tau$ specifies the proton and neutron contributions to the generalized rearrangement
terms with coefficients ${\cal{R}}^{\tau}_{mn}$. It is the presence of the rearrangement terms that transforms Eq. (\ref{eq4})
into a non-linear eigenvalue problem. As can be seen from Eq. (\ref{e889}), the rearrangement terms contain contributions
associated with the one-body density $\rho$ and the two-body correlation matrix $\sigma$ defined by
\begin{equation}
\dspt \sigma_{ij,kl}= \langle \Psi \vert a_{i}^{+} a_{k}^{+} a_{l} a_{j} \vert \Psi  \rangle  - \rho_{ji} \rho_{lk}
+ \rho_{jk} \rho_{li}.
\end{equation}
In Appendix \ref{a1}, we express Eqs. (\ref{eq4})-(\ref{eq5}) in a "core + valence space" scheme, a form which is explicitly
used in the present study.

The minimization of ${\cal{F}}(\rho)$ with respect to the single-particle orbitals leads to inhomogeneous Hartree-Fock (HF)
equations which depend on the amount of correlations contained in $\vert \Psi \rangle$ \cite{pillet},
\begin{equation}
[h(\rho, \sigma), \rho] = G(\sigma),
\label{eqad1}
\end{equation}
where
\begin{equation}{
\begin{array}{l}
\dspt G_{kl}(\sigma) = \frac{1}{2} \sum_{imn} \langle im \vert V(\rho) \vert kn \rangle \sigma_{il,mn}  \\
\dspt ~~~~~~~~~-\frac{1}{2} \sum_{imn} \langle ml \vert V(\rho) \vert \widetilde{ni} \rangle \sigma_{ki,mn}
\end{array}}
\label{ia51}
\end{equation}
In Eq.(\ref{eqad1}), $h(\rho, \sigma)$ is the one-body mean-field Hamiltonian built with the one-body density $\rho$ and the
two-body correlation matrix $\sigma$:
\begin{equation}
h_{ij}(\rho, \sigma) = \langle i \vert K \vert j \rangle + \Gamma_{ij}(\rho) + \partial \Gamma_{ij}(\rho) + \partial \Gamma_{ij}(\sigma).
\label{eqad2}
\end{equation}
Explicit expressions for the fields $\Gamma_{ij}(\rho)$, $\partial \Gamma_{ij}(\rho) $ and $\partial \Gamma_{ij}(\sigma)$ are
given in Ref. \cite{pillet}. The basis that diagonalizes $h(\rho, \sigma)$ provides proton and neutron single-particle orbitals
with energies $\epsilon_{i}^{\tau}$. The diagonal part of Eq. (\ref{eq5}), that is the one obtained by taking
$\alpha \equiv \alpha'$, can be easily written in terms of the $\epsilon_{i}^{\tau}$.
In the limit where the mixing reduces to the sole HF configuration, one recovers the standard HF expression for
$\epsilon_{i}^{\tau}$,
\begin{equation}
\begin{array}{c}
\dspt \epsilon_{i}^{\tau} = \langle i_{\tau} \vert K \vert i_{\tau} \rangle +
\sum_{\tau'} \sum_{h=1}^{N^{\tau'}} \langle i_{\tau} h_{\tau'} \vert V(\rho) \vert i_{\tau} h_{\tau'} \rangle  \\
\dspt + \frac{1}{2} \sum_{\tau' \tau''} \sum_{h'=1}^{N^{\tau'}} \sum_{h''=1}^{N^{\tau''}}
\langle h'_{\tau} h''_{\tau''} \vert \frac{\partial V(\rho)}{\partial \rho_{i_{\tau} i_{\tau}}}
\vert h'_{\tau'} h''_{\tau''}\rangle.
\end{array}
\label{eqqq6}
\end{equation}
In Eq. (\ref{eqqq6}), summations over $h, h'$ and $h''$ run over the hole states.

\begin{figure*}[htb!]
\begin{center}
\includegraphics[height=5.0cm]{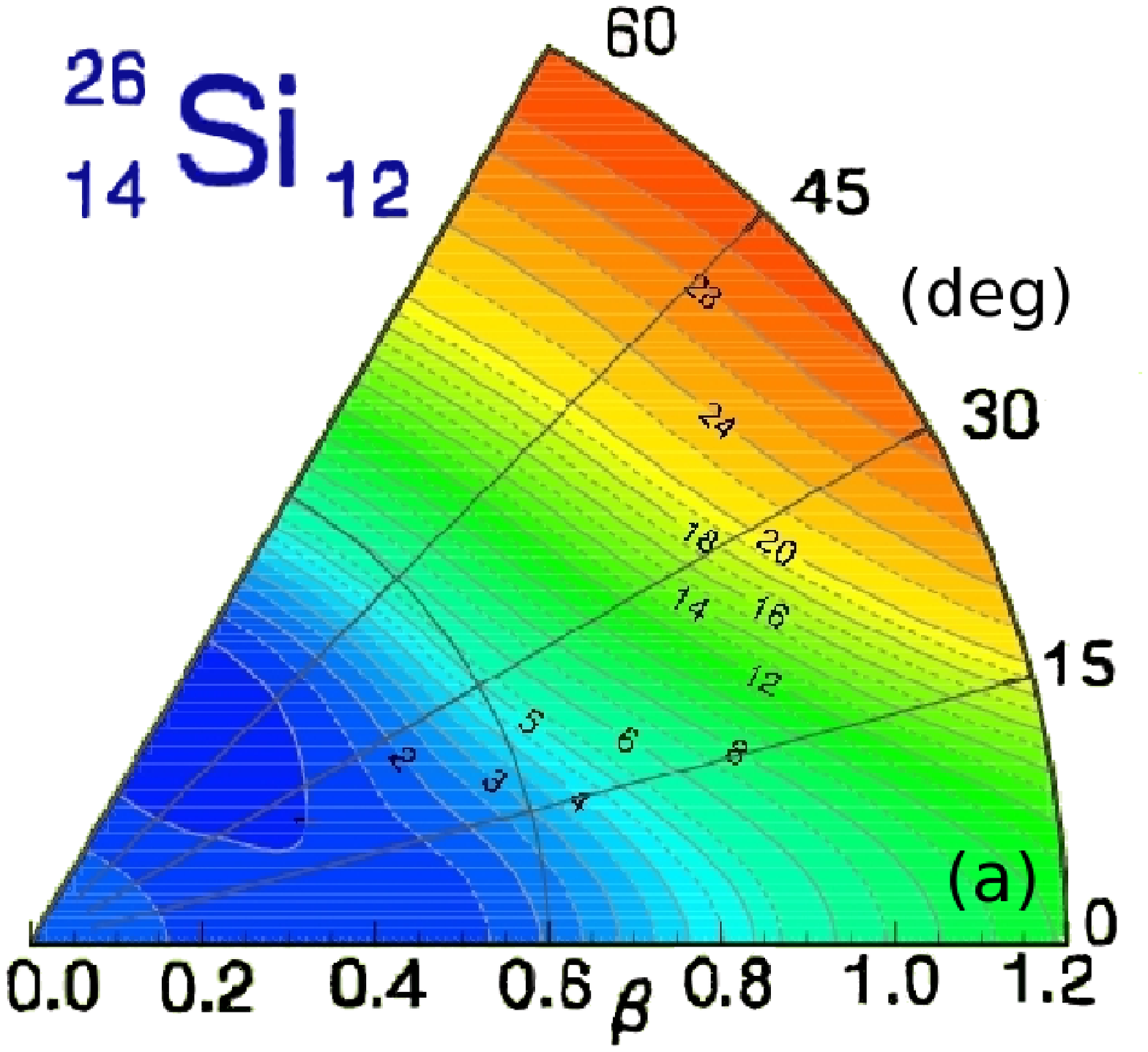} \hspace{2.5cm}
\includegraphics[height=5.0cm]{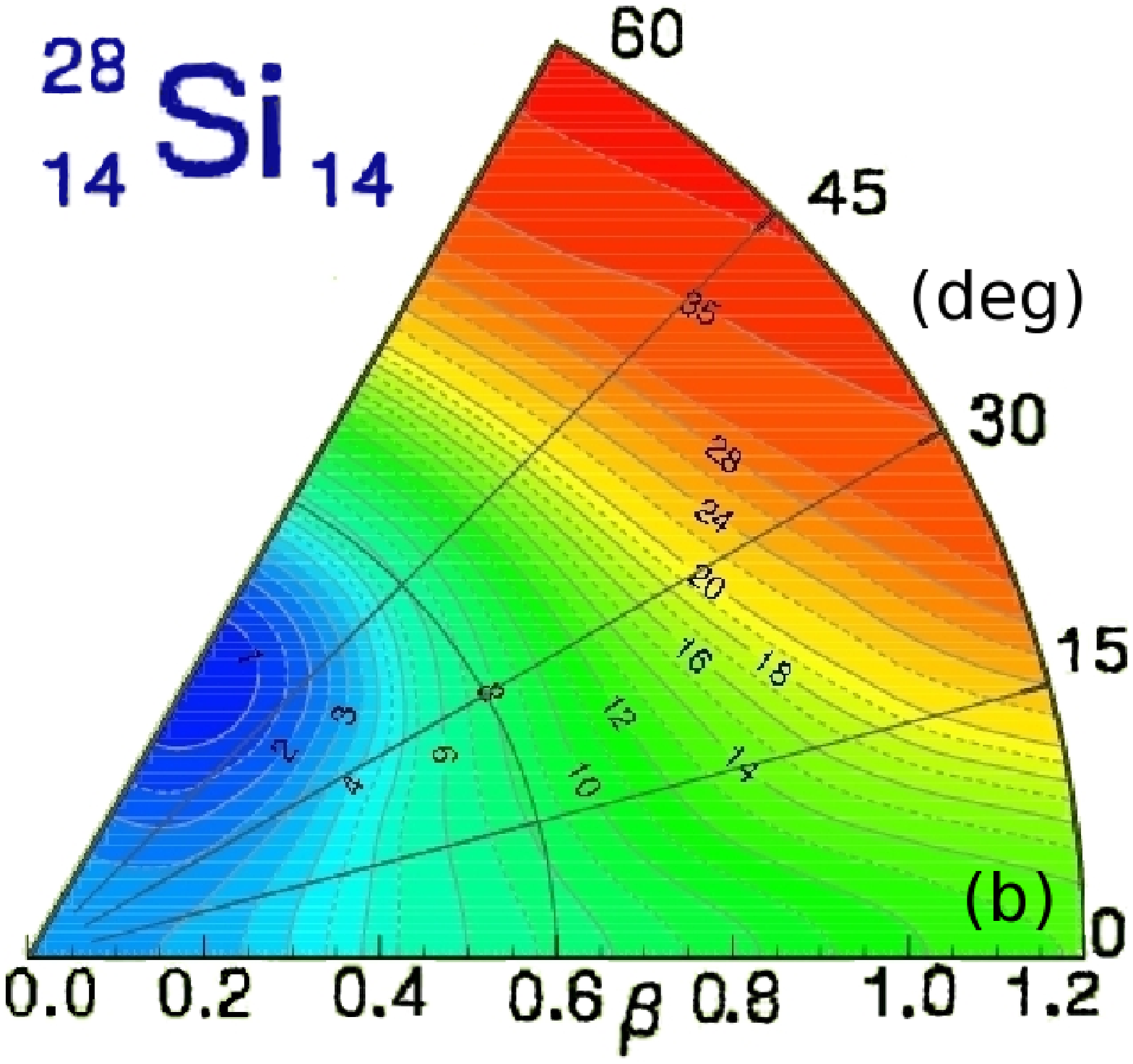}
\vspace{0.4cm}
\includegraphics[height=5.0cm]{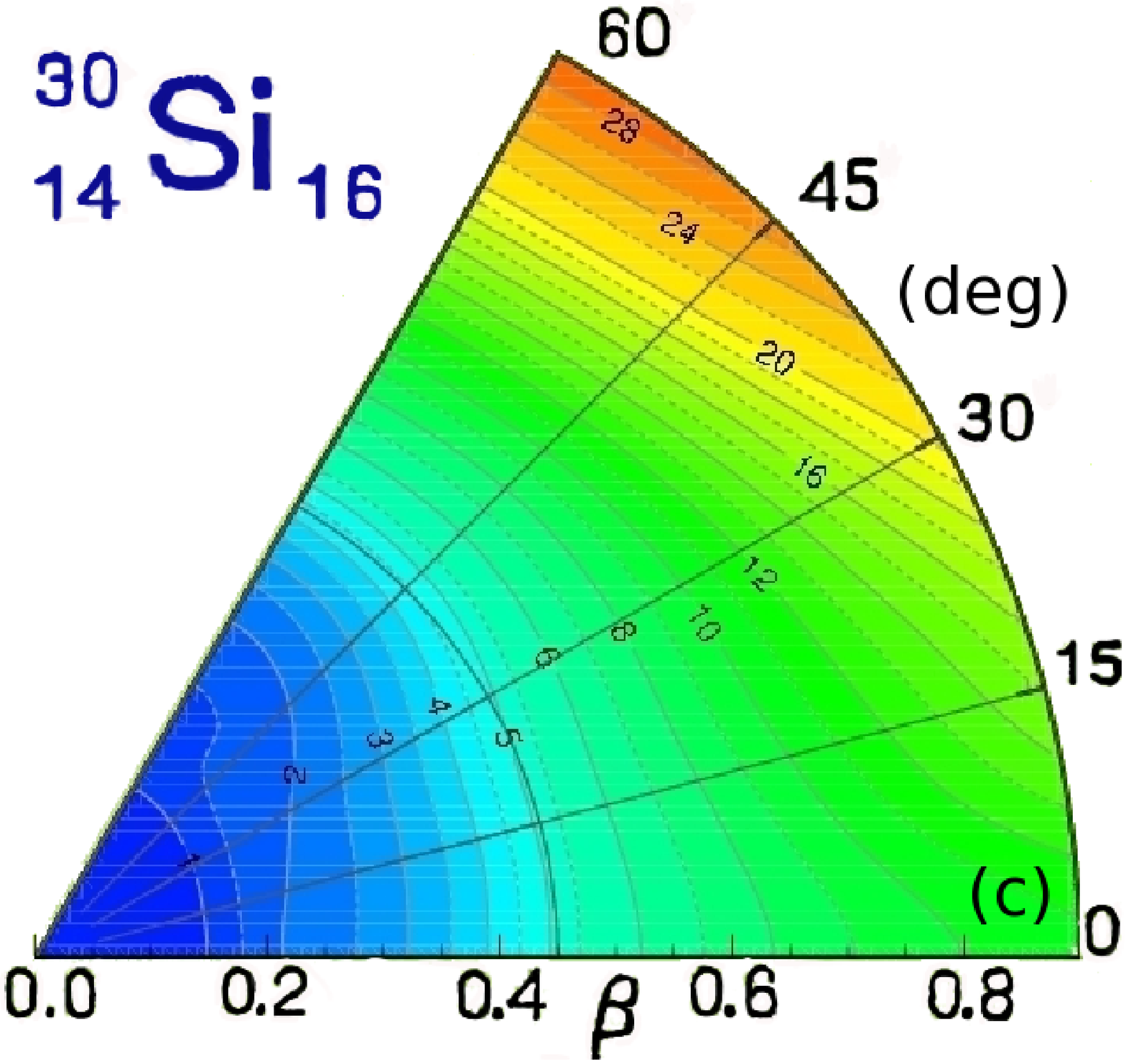} \hspace{2.5cm}
\includegraphics[height=5.0cm]{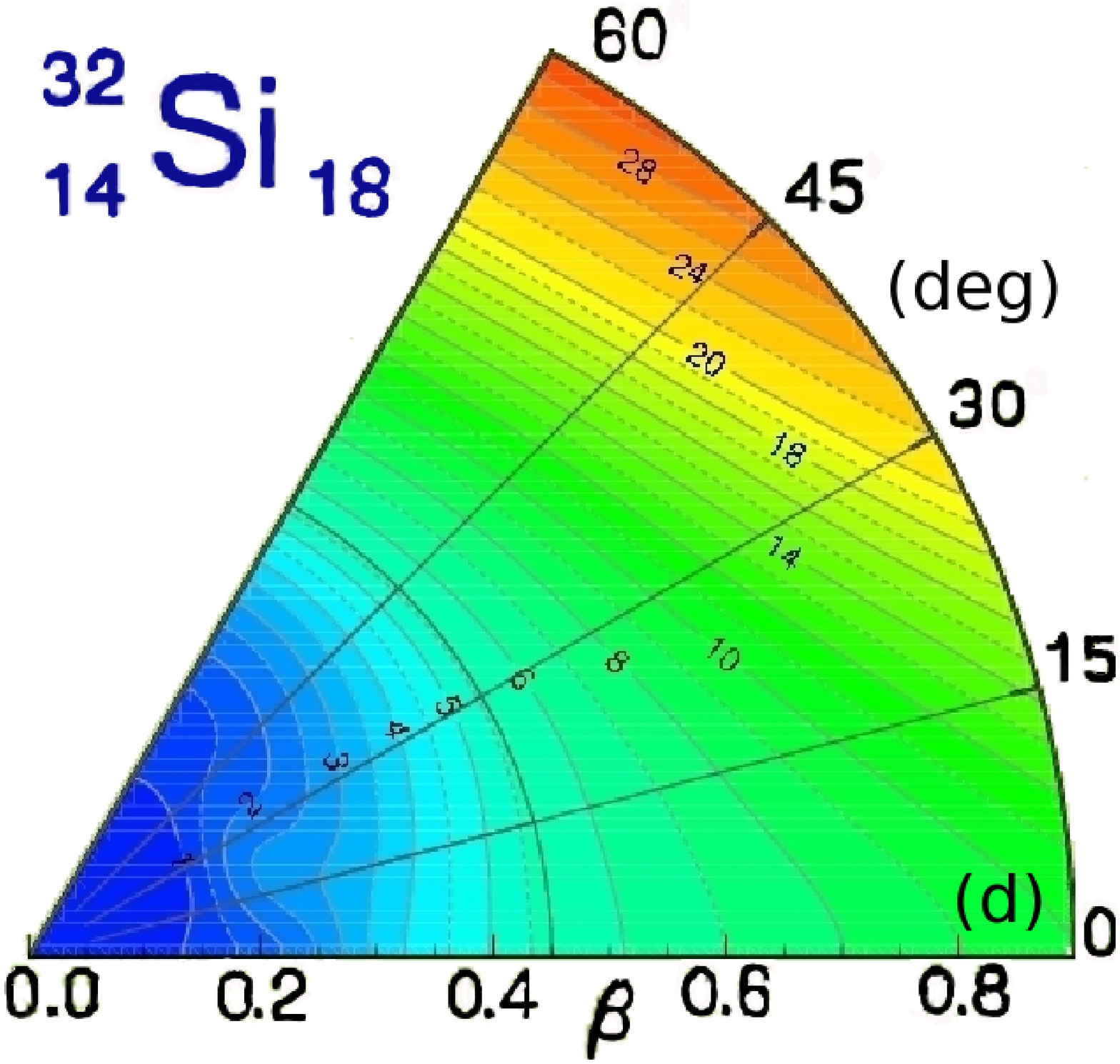}
\end{center}
\caption{(Color online) Triaxial HFB potential energy surfaces for $^{26-32}$Si (panels (a), (b),
(c) and (d), respectively) obtained with the D1S Gogny interaction.}
\label{fig1}
\end{figure*}

As discussed in Ref.\cite{pillet}, a fully self-consistent
solution of the multiconfiguration approach
consists of solving simultaneously both equations (\ref{eq4}) and (\ref{eqad1}).
It is important to notice that,
in our approach, the mean field and beyond mean-field descriptions are obtained
in a consistent way since they both follow from variations of the same
functional. A standard way to solve both equations is to use an iterative
procedure. Starting from a HF solution (characterized by $\rho=\rho_{HF}$ and
$\sigma=0$) that provides an initial set of single particle orbitals,
one builds the multiconfiguration basis and solves the configuration mixing
Eq. (\ref{eq4}). Then, from
the mixing coefficients obtained from
Eq. (\ref{eq4}), one calculates $\rho$ and $\sigma$ and solves Eq. (\ref{eqad1}).
With the new set of single particle orbitals, one redoes the procedure until
convergence of $\rho$ and $\sigma$.
In this way, the single-particle orbitals
contain effects coming not only from the mean-field built with the one-body
density matrix $\rho$ but also from the correlations contained in the matrix
$\sigma$.

In general, the actual solution of Eq. (\ref{eqad1}) is a very difficult task.
In a previous study dedicated to the particular case of pairing-type
correlations in the ground states of even-even tin isotopes \cite{pillet}, one
has solved Eq. (\ref{eqad1}) in an approximate way, by neglecting $\sigma$. In
the present study of even-even silicon isotopes, as a first step, we have solved only
the configuration interaction (CI) part (Eq. (\ref{eq4})) of the
multiconfiguration approach.

As already pointed out in Introduction,
multiconfiguration methods are able to preserve certain symmetries. Concerning
particle numbers, the correlated wave function (\ref{eq2})
is a superposition of Slater determinants
that conserves exactly the numbers of protons and neutrons.
As to angular momentum, the multiconfiguration equations have
been solved in the present study only for spherical nuclear configurations,
with valence spaces comprising complete spherical subshells.
As a consequence, nuclear states have a good total angular momentum $J$.
Actually, since our formalism has been developed in axial
symmetry in order to be able to introduce
quadrupole deformation explicitly, the only
conserved quantum numbers we have are the projection $K$ of the
total angular momentum and the parity.
The conserved quantum number $J$ is then obtained in the usual
manner by performing calculations for successive values $K=0,1,2,\ldots$ and
identifying degenerate multiplets. Let us note that the finite axial harmonic
oscillator bases used for expanding single-particle states (see below) are
defined with a spherical truncation.

\section{Low-lying states in $^{26-32}$Si}\label{section2}

This section is devoted to the description of the low-lying spectroscopy of $^{26-32}$Si using the $mp-mh$ configuration mixing approach. In the first part of this analysis, we investigate the mean-field properties of the ground states provided by the Hartree-Fock-Bogoliubov (HFB) approach using the same D1S Gogny interaction and compare them with the results of a five-dimensional approximate GCM approach.

Technically, the single-particle states introduced
in the different approaches (HF, HFB, $mp-mh$)
are expanded onto the harmonic oscillator (HO) bases. In
the present work,
$N_{0}$=11 major spherical shells have been taken.
This basis size has been found
sufficient to ensure the convergence of all the results obtained in this work in
the three approaches mentioned above. For instance, the convergence of low-lying state energies in the $mp-mh$ configuration mixing approach has been achieved within an accuracy better than 0.1 keV.

\subsection{Ground state deformation properties}\label{section2A}

In order to investigate the properties of the ground states of $^{26-32}$Si, we start with triaxial HFB calculations
constrained according to the dimensionless deformation parameters $\beta$ and $\gamma$,
\begin{equation}
\beta = \frac{\sqrt{5 \pi} \sqrt{q_{0}^{2}+3q_{2}^{2}}}{3A^{5/3} r_{0}^{2}} \hspace{0.5cm} {\rm and}
\hspace{0.5cm} \gamma =\tan^{-1}\left( \sqrt{3}\,\frac{q_{2}}{q_{0}}\right).
\label{eqad4}
\end{equation}
In Eq. (\ref{eqad4}), $q_{0} = \langle \hat{Q}_{0} \rangle = \langle 2z^2-x^2-y^2 \rangle$,
$q_{2} = \langle \hat{Q}_{2} \rangle = \langle x^2-y^2 \rangle$ and $r_{0}$=1.2 fm.

Fig. \ref{fig1} displays the triaxial HFB potential energy surfaces (PES) of $^{26-32}$Si in the $\beta$ and $\gamma$
degrees of freedom. Isolines associated with total energy are indicated with a numbering corresponding
to the height of the potential (in MeV) relative to the minimum of the HFB potential (dark blue online) for each nucleus.
The $\beta_{{\rm HFB}}$ and $\gamma_{{\rm HFB}}$ values for the HFB solution of lowest energy
are indicated in the second and the third columns of Table \ref{tab6d}.

\begin{table}[htb!]
\begin{center}
\begin{tabular}{ccccc}
\hline
\hline
  Nucleus   &  $\beta_{{\rm HFB}}$  & $\gamma_{{\rm HFB}}$ &  $\langle \beta \rangle_{5{\rm DCH}}$  &  $\langle \gamma \rangle_{5{\rm DCH}}$  \\
\hline
 $^{26}$Si  &    0.32      & 60.0$^{\circ}$    &  0.41      &  28$^{\circ}$  \\
 $^{28}$Si  &    0.37      & 60.0$^{\circ}$    &  0.40      &  27$^{\circ}$  \\
 $^{30}$Si  &    0.00      & 0.0$^{\circ}$      &  0.39      &  29$^{\circ}$  \\
 $^{32}$Si  &    0.01      & 34.0$^{\circ}$    &  0.37      &  28$^{\circ}$  \\
\hline
\hline
\end{tabular}
\end{center}
\caption{ $\beta$ and $\gamma$ deformation properties of the $^{26-32}$Si ground states
from the HFB and 5DCH approaches and the D1S Gogny interaction.}
\label{tab6d}
\end{table}

From Table \ref{tab6d}, one sees that the HFB ground states of $^{26}$Si and $^{28}$Si are
similarly characterized by a large value of $\beta_{{\rm HFB}}$ $\simeq$ 0.35 and $\gamma_{{\rm HFB}}=60^{\circ}$.
The HFB minima are found well-deformed on the oblate side. $^{30}$Si and $^{32}$Si exhibit different characteristics.
$\beta_{{\rm HFB}}$ is equal to zero for $^{30}$Si and very close to zero for $^{32}$Si.
The values of $\gamma_{{\rm HFB}}$ are very different, $\gamma_{{\rm HFB}}=0^{\circ}$ for $^{30}$Si and $\gamma_{{\rm HFB}}=34^{\circ}$ for $^{32}$Si. They can be considered
as spherical and nearly spherical nuclei, respectively.

Investigating the PESs of Fig. \ref{fig1}, one can deduce that, even though $\beta_{{\rm HFB}}$ and $\gamma_{{\rm HFB}}$ may be quite different in the four nuclei, the common feature of the four PESs is their softness in both the $\beta$ and $\gamma$ degrees of freedom.
In relative, the $^{26}$Si PES appears to be more $\gamma-$soft than the one of $^{28}$Si.
The PESs of $^{30}$Si displays similar features as the one of $^{32}$Si.
This proposes an important role of triaxiality.
This softness is quantitatively evidenced by the mean values $\langle \beta \rangle_{5{\rm DCH}}$ and $\langle \gamma \rangle_{5{\rm DCH}}$ displayed in Table \ref{tab6d}. These values have been obtained from a five-dimensional
collective Hamiltonian (5DCH) describing both $\beta - \gamma$ and rotation modes with the use of the approach developed in Ref. \cite{delaroche,libert}.
Let us recall that such an approach is based on the completely microscopic generator coordinate method (GCM) \cite{ring}
and allows one to find collective excitations of pure rotational-vibrational character from the only data of the nucleon-nucleon
effective force.
The 5DCH calculations predict strong dynamical $\beta$-deformations for the ground states of all four silicon isotopes together with significant triaxiality. One observes that the 5DCH collective dynamics introduces considerable changes with respect to HFB in the $\beta - \gamma$ ground state deformations.

\begin{figure*}[htb!]
\begin{center}
\includegraphics[height=5.0cm]{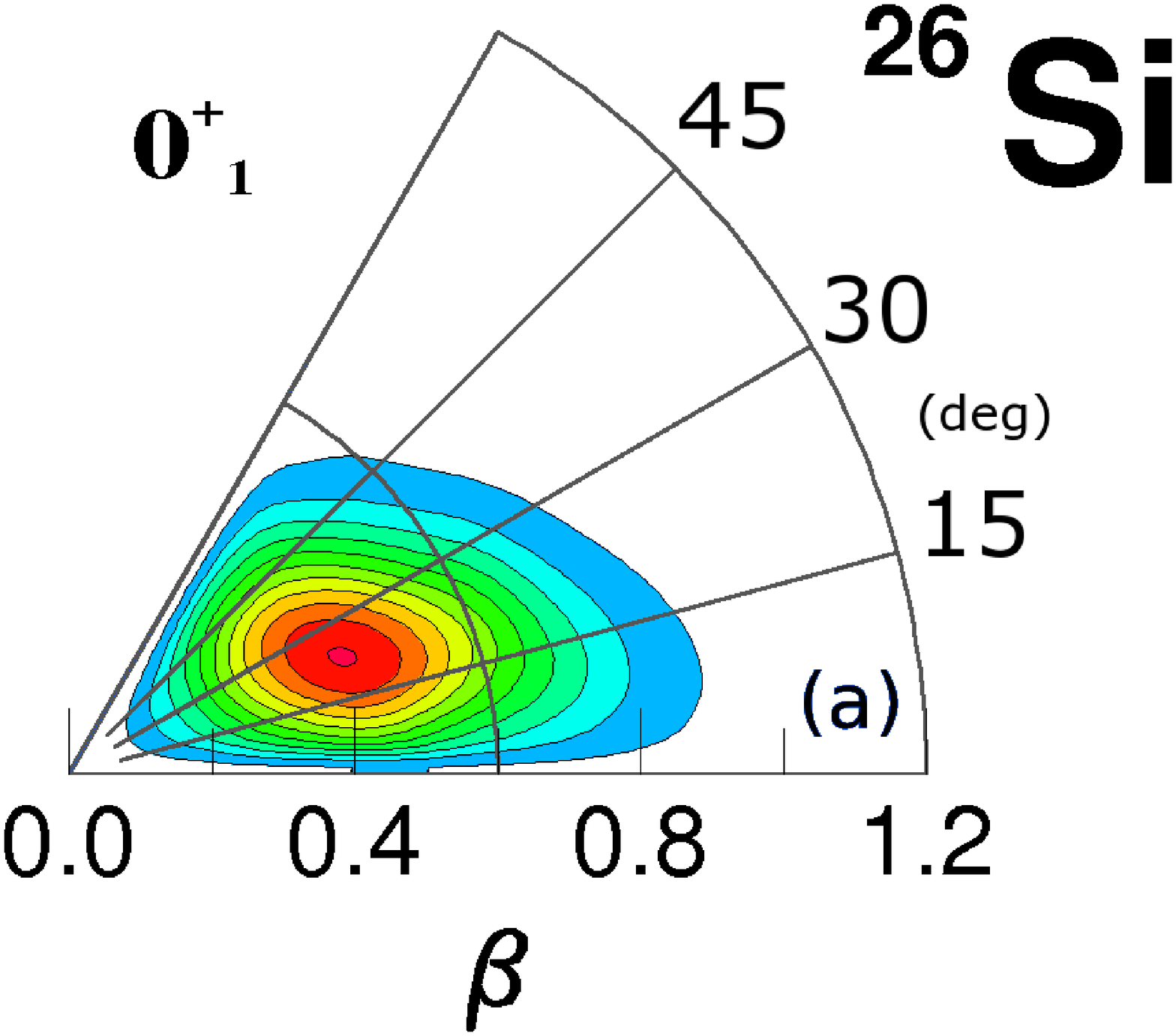} \hspace{2.5cm}
\includegraphics[height=5.0cm]{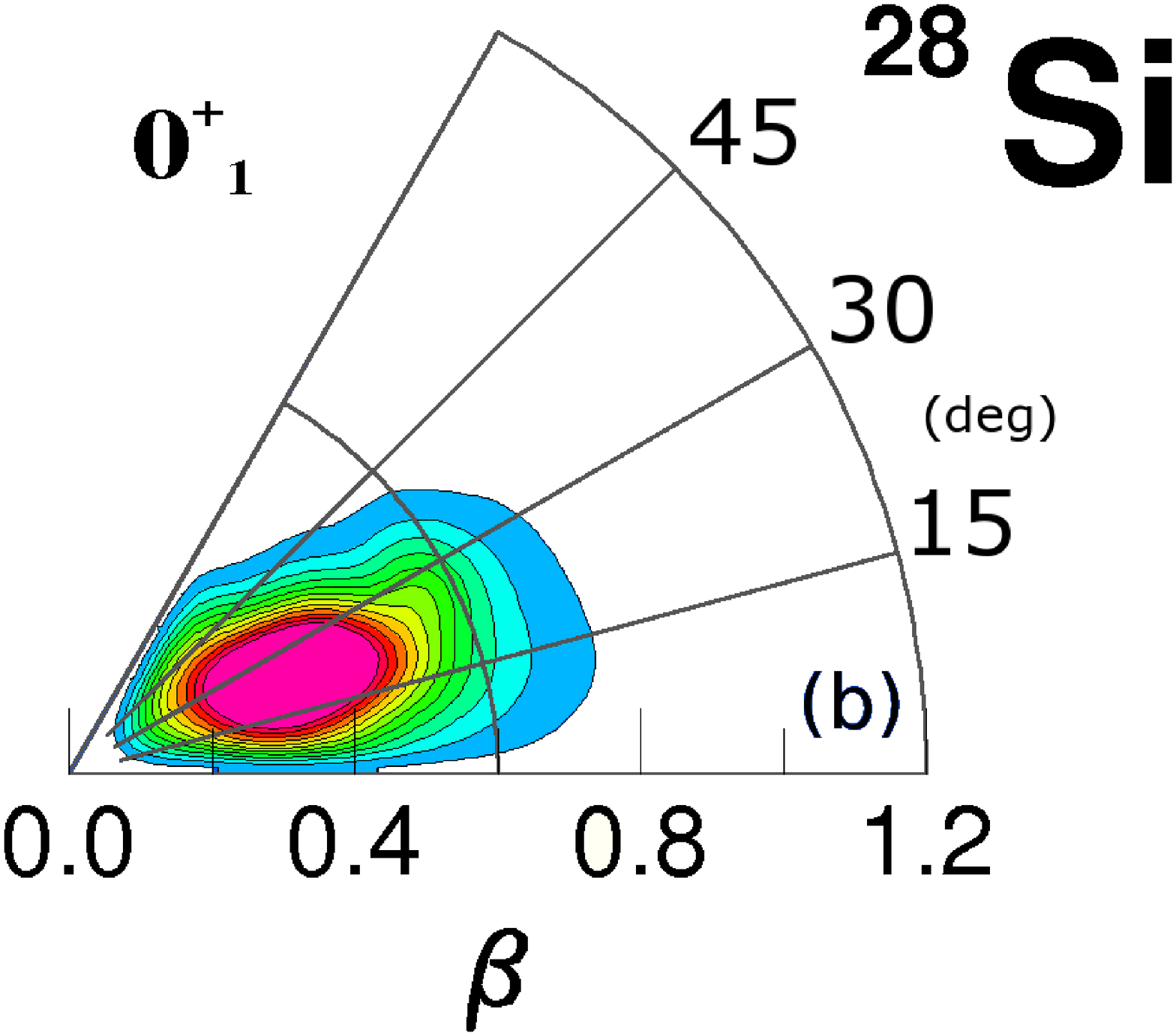}
\vspace{1.0cm}
\includegraphics[height=5.0cm]{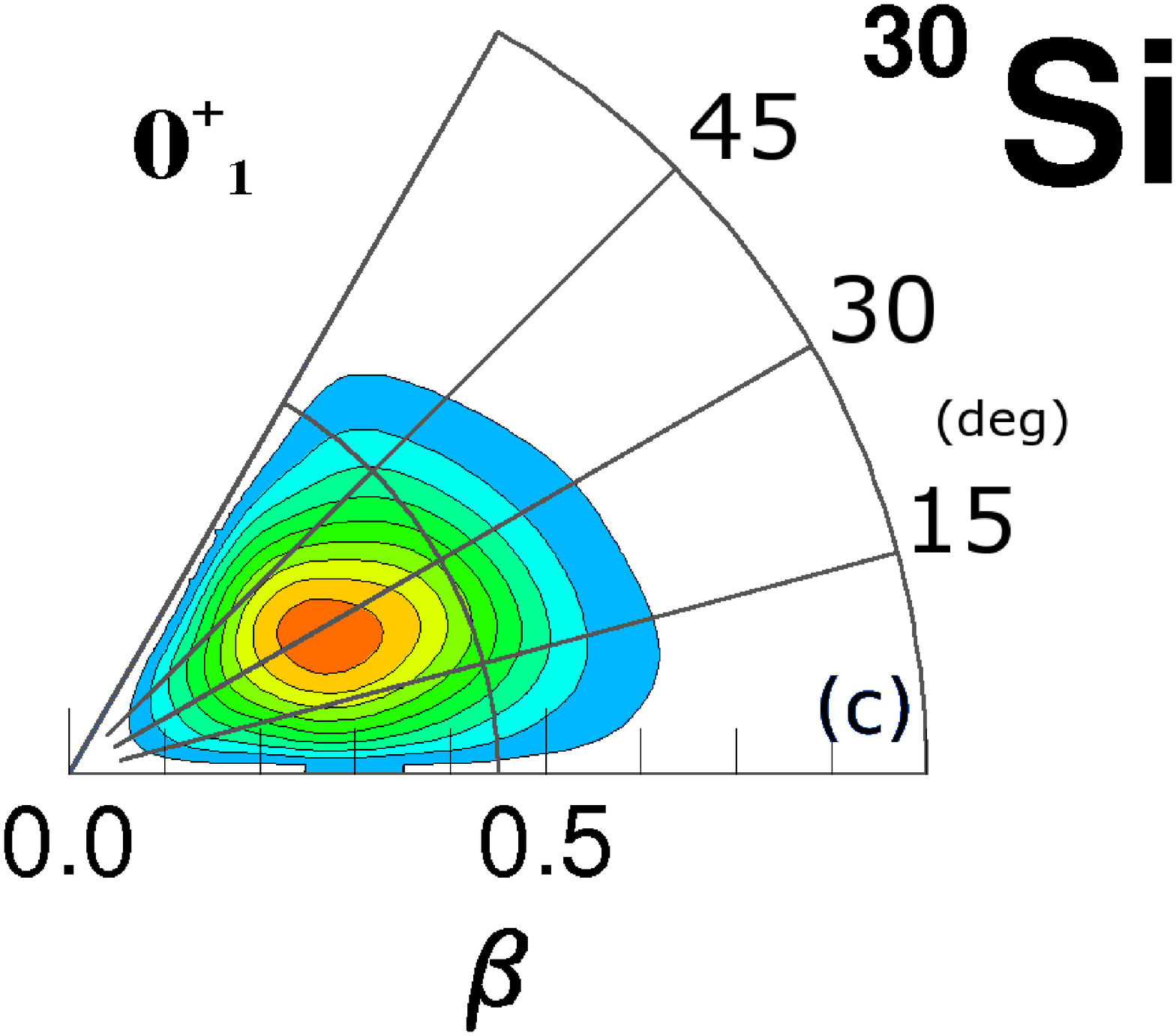} \hspace{2.5cm}
\includegraphics[height=5.0cm]{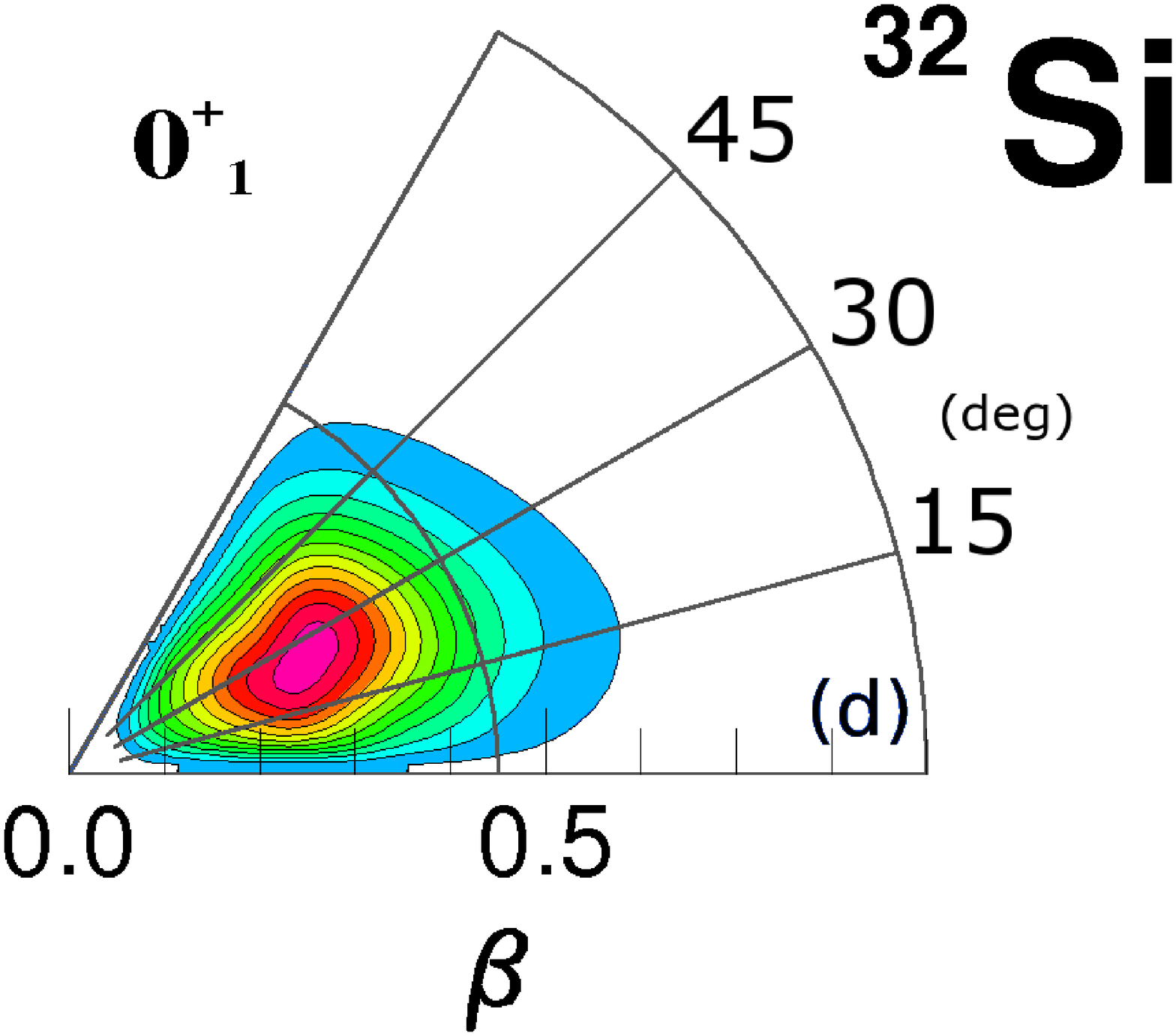}
\end{center}
\caption{(Color online) Ground state 5DCH collective wave functions for $^{26-32}$Si (panels (a), (b),
(c) and (d), respectively).}
\label{figfdo}
\end{figure*}

To detail the information provided by $\langle \beta \rangle_{5{\rm DCH}}$ and $\langle \gamma \rangle_{5{\rm DCH}}$ mean values, Fig. \ref{figfdo} shows the 5DCH collective wave
functions of  $^{26-32}$Si ground states (panels (a), (b), (c) and (d), respectively).
Even though the $^{28}$Si collective wave function is partly suppressed for $\gamma$ between
45 and 60 degrees, the spreading of collective wave functions in $\beta - \gamma$ plane is evident. The spreading in $\beta$ goes up to $\sim$0.8 for $^{26}$Si and $^{28}$Si and
$\sim$0.6 for $^{30}$Si and $^{32}$Si.

\begin{table}[h!]
\begin{center}
\begin{tabular}{ccccccccc}
\hline
\hline
  & $\pi$d$_{5/2}$ & $\pi$s$_{1/2}$ & $\pi$d$_{3/2}$ & $\nu$p$_{1/2}$  & $\nu$d$_{5/2}$ & $\nu$s$_{1/2}$ & $\nu$d$_{3/2}$ & $\nu$f$_{7/2}$ \\
\hline
$^{26}$Si& -7.07  & -2.61  &  0.88 &-27.73 & -15.96 & -10.73 & -7.33 & -1.37 \\
$^{28}$Si& -10.05 & -5.16  & -1.69 &-28.39 & -15.95 & -10.91 & -7.34 & -1.18 \\
$^{30}$Si& -12.60 & -7.32  & -4.11 &-26.75 & -16.21 & -11.56 & -7.29 & -1.41 \\
$^{32}$Si& -15.09 & -10.08 & -7.51 &-26.44 & -16.28 & -11.40 & -7.80 & -2.01 \\
\hline
\hline
\end{tabular}
\end{center}
\caption{Energies (in MeV) of spherical proton and neutron HFB single-particle states in $^{26-32}$Si.}
\label{tab4}
\end{table}

In Table \ref{tab4} we report the energies $\epsilon_{i}^{\tau}$ of the spherical orbitals for $^{26-32}$Si, for protons
($\pi 1d_{5/2}$, $\pi 2s_{1/2}$, $\pi 1d_{3/2}$) and neutrons ($\nu 1p_{1/2}$, $\nu 1d_{5/2}$, $\nu 2s_{1/2}$,
$\nu 1d_{3/2}$, $\nu 1f_{7/2}$).
These energies slowly evolve from one isotope to the other. We notice a slight increase of the neutron gap between
$\nu 2s_{1/2}$ and $\nu 1d_{3/2}$ orbitals in
$^{30}$Si, which reaches a maximum value of $\simeq\,4.56$ MeV. A previous study of $N=16$ isotones with the
D1S Gogny interaction have suggested that all $Z=10-18$ isotones show strong deformations, limiting the understanding
of $N=16$ as a magic number to the sole oxygen neutron drip line \cite{obertelli}.

\begin{figure}[h!]
\begin{center}
\includegraphics[height=8.0cm]{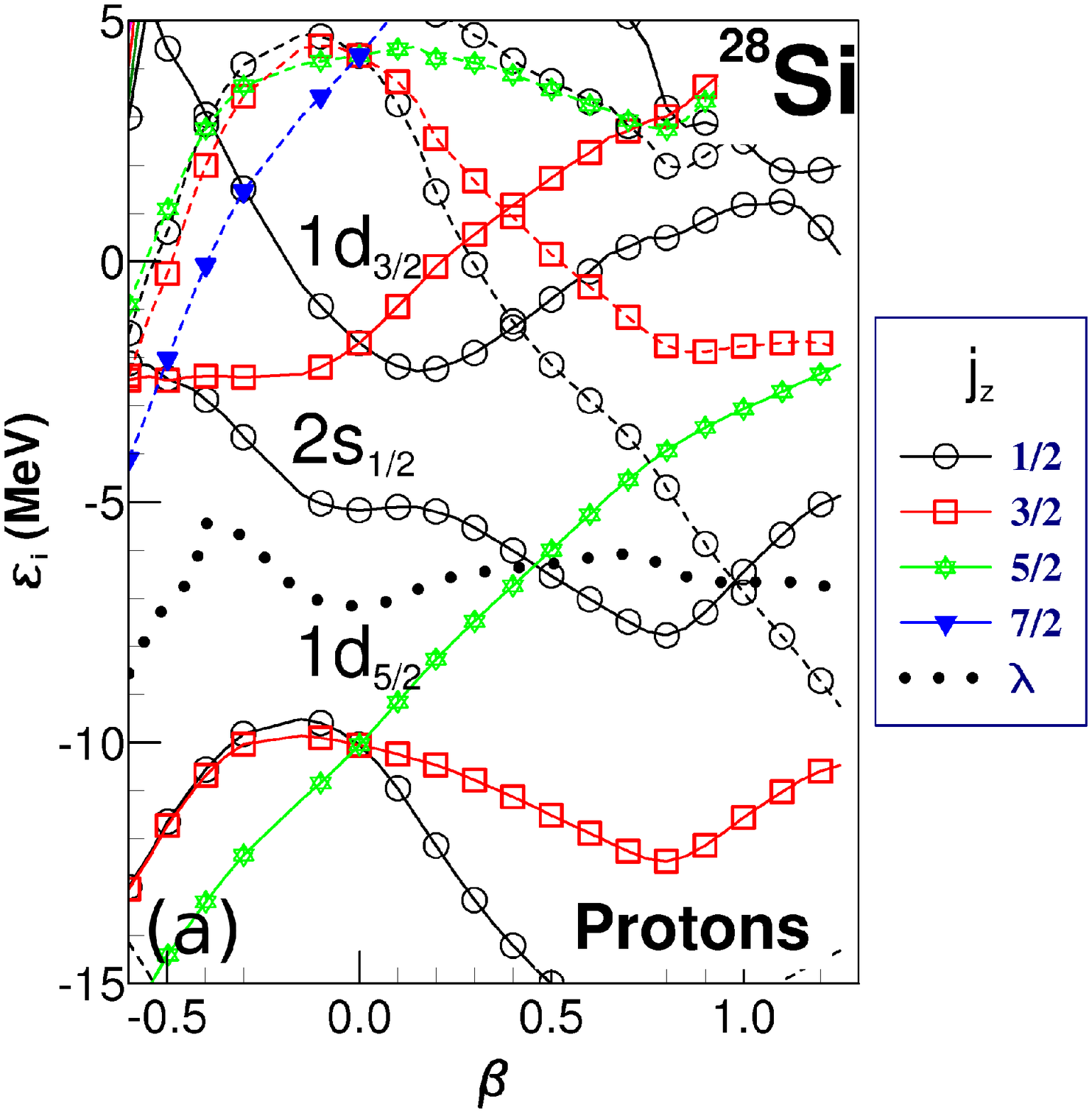}
\includegraphics[height=8.0cm]{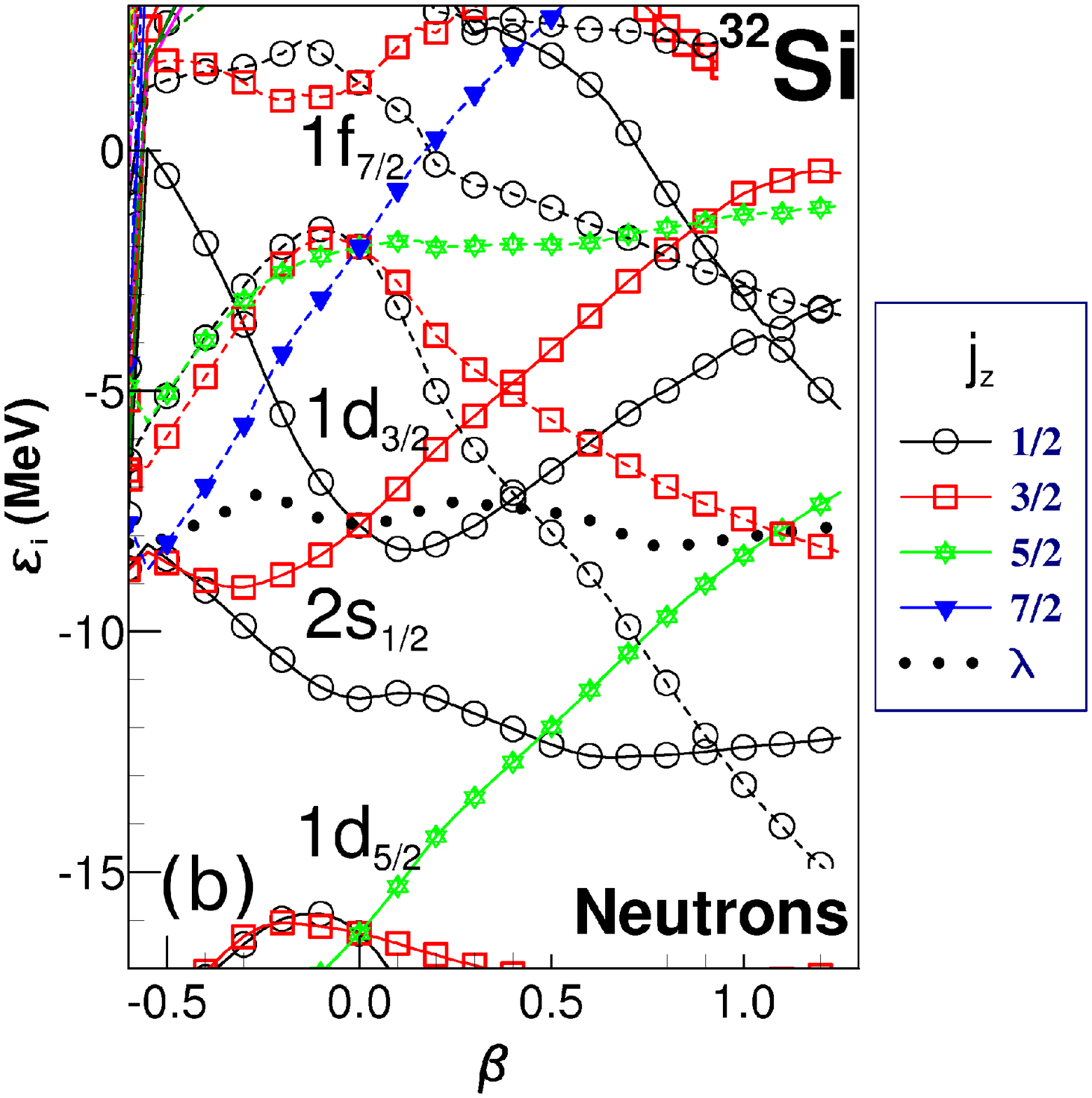}
\end{center}
\caption{(Color online) Evolution of proton and neutron single-particle orbitals obtained
within the HFB approximation
with the axial deformation $\beta$ ($\gamma=0^{\circ}$) in $^{28}$Si (upper panel (a)) and $^{32}$Si (lower panel (b)).}
\label{fig1b}
\end{figure}

The evolution of proton and neutron single-particle orbitals obtained within the HFB approximation with
axial deformation $\beta$ ($\gamma=0^{\circ}$)
is displayed in Fig. \ref{fig1b}. To discuss the general characteristics of single-particle orbitals
when spherical symmetry is broken, we have arbitrarily selected the isotopes $^{28}$Si (for protons) and $^{32}$Si 
(for neutrons). Open circles (black online), squares (red online), stars (green online) and triangles (blue online) stand
for the projections of the angular momentum on the symmetry axis with values $j_{z}=1/2, ~3/2, ~5/2$
and $7/2$, respectively. Solid (dashed) lines correspond to positive (negative) parity orbitals. The chemical potential is indicated
with filled black circles and is denoted by $\lambda$.

For protons (upper panel), starting from the oblate side and up to the spherical shape,
the Fermi level corresponds to the $j_z$=1/2 deformed orbital originating from the $1d_{5/2}$ shell. From
sphericity up to a prolate deformation $\beta \simeq$0.5, the Fermi level is located on the $j_z$=5/2 deformed
orbital and then migrates to the $j_z$=1/2 orbital coming from the
$2s_{1/2}$ shell; finally, for a very large value of $\beta $, it follows the $j_z$=1/2 deformed orbital coming from
the $1f_{7/2}$ shell. The $j_z$=1/2 and $j_z$=3/2 deformed orbitals from the $1d_{5/2}$ shell for small oblate
deformations fall down instead of going up. Investigating this plot in more detail, one sees
that similar trends are encountered for deformed orbitals originated from the $2s_{1/2}$
and $1d_{3/2}$ shells. One is led to the conclusion that
these orbitals are strongly mixed through deformation as confirmed by the presence of avoided level crossings
in single-particle spectra. The natural continuity of
the oblate $j_z$=1/2 from the $1d_{5/2}$ shell is the prolate $j_z$=1/2 from the $1d_{3/2}$ shell. Other examples
are the oblate $j_z$=1/2 from the $2s_{1/2}$ shell and the prolate $j_z$=1/2 from the $1d_{5/2}$ shell,
the oblate $j_z$=1/2 level from the $1d_{3/2}$ shell and the prolate $j_z$=1/2 from the $2s_{1/2}$ shell, or
the oblate/prolate $j_z$=3/2 from the $1d_{5/2}$ shell and the prolate/oblate $j_z$=3/2 from the $1d_{3/2}$ shell.
From this kind of analysis, one expects that the proton excitations contributing to low-lying states will
mainly arise from the $sd$-shell, while the influence of the $j_z$=1/2 level from the 1$f_{7/2}$ shell will appear only at quite
large $\beta$ deformation.
It is important to note that these mixings and repulsions of deformed orbitals are not necessarily synonymous with
inversions of orbitals.

The behavior of the neutron orbitals is similar. As in the present paper we study isotopes where the
$2s_{1/2}$ and $1d_{3/2}$ neutron subshells are filled or
partially filled, it is important to look at higher shells. Here we only mention that the same game as for protons is
played between deformed neutron orbitals originating from $1f_{7/2}$, $2p_{3/2}$ and $2p_{1/2}$ with
low $j_z$ values, see Fig. \ref{fig1b}, lower panel, and that upper shells of negative parity play a role at small
deformations only for the heavier isotopes.

\subsection{Low-lying spectroscopy}\label{lls}

In this subsection we discuss the low-lying states in the even-even $^{26-32}$Si isotopes as predicted by the
$mp-mh$ configuration mixing approach and compare them with the ones obtained with the 5DCH method.
Only positive parity states are investigated in this work.
Experimentally, these nuclei are challenging as they exhibit a large variety of states at low energy,
and strong changes arise from one isotope
to another (for example, in the 2$^{+}_{2}$ state). For this reason, from the theoretical viewpoint, they can be considered as
benchmarks for both the many-body method employed and the properties of the effective nucleon-nucleon interaction used.
The comparison with the 5DCH approach, a method that has proved its pertinence through a global survey \cite{delaroche},
is interesting in the sense that, as the same D1S Gogny interaction is used, it enables one to specify the role of
rotational and quadrupole correlations in the spectroscopy of the positive parity states in silicon isotopes.

All the $mp-mh$ results we present below correspond to mixing within the $sd$-shell.
We have checked the influence on the low-lying states of our interest when including the $1p$ and $1f$ subshells.
In particular, the adding of $1p_{1/2}$ and $1p_{3/2}$ had essentially no effect on the spectroscopy of
$^{26}$Si, $^{28}$Si, $^{30}$Si and $^{32}$Si.
For the selected spectroscopy of $^{26}$Si, the effect of the $1f_{7/2}$ shell was very small. For $^{28}$Si, only small
variations were observed and they concerned only the highest states. The largest differences ($\sim$ 500 keV) due to the 1$f_{7/2}$ and 2$p_{3/2}$ orbitals have been encountered in $^{30}$Si and $^{32}$Si.
\begin{table}[h!]
\begin{center}
\begin{tabular}{ccccc}
\hline
\hline
  Nucleus   &  States        & Experiment &   $mp-mh$    & 5DCH    \\
\hline
 $^{26}$Si  &                &            &           &         \\
            & 2$^{+}_{1}$    &   1.795    &  1.502    &  2.426  \\
            & 2$^{+}_{2}$    &   2.783    &  2.567    &  5.124  \\
            & 0$^{+}_{2}$    &   3.332    &  3.740    &  8.146  \\
            & 3$^{+}_{*}$    &   3.756    &  3.233    &  9.126  \\
            & 4$^{+}_{*}$    &   3.842    &  3.293    &  6.119  \\
            & 3$^{+}_{*}$    &   4.093    &  3.779    &    \\
            & 2$^{+}_{3}$    &   4.138    &  3.915    &   \\
            & 4$^{+}_{*}$    &   4.183    &  4.758    &  9.254  \\
            & 2$^{+}_{*}$    &   4.446    &  4.783    &    \\
            & 0$^{+}_{3}$    &   4.806    &  4.959    &    \\
            & 4$^{+}_{*}$    &   5.229    &  4.931    &    \\
            & 4$^{+}_{1}$    &   5.330    &  5.741    &    \\
            & 2$^{+}_{*}$    &   5.562    &  5.577    &    \\
            & 3$^{+}_{?}$    &     ?      &  5.755    &    \\
            & 0$^{+}_{4}$    &   5.940    &  6.690    &    \\
            & 2$^{+}_{4}$    &   6.350    &  6.823    &    \\
\hline
\hline
\end{tabular}
\end{center}
\caption{Excitation energies (in MeV) of positive parity low-lying states in $^{26}$Si
from experiments (second column), and calculated with the variational $mp-mh$ configuration mixing method using
the D1S Gogny force (third column). In the fourth column a few energies derived from the 5DCH approach \cite{delaroche}
are displayed. The symbol * means that the spin of the state is not experimentally assigned.
The symbol ? indicates a state which is not seen experimentally.}
\label{tab5a}
\end{table}
Concerning the 5DCH approach, only theoretical results with energies lower than $\simeq$12 MeV are presented. This choice
is not fully arbitrary as it is motivated by the relative maximum height of
PESs ($\simeq\,$28 MeV above the HFB minimum in the present case, see Fig. \ref{fig1}) used to perform the GCM configuration
mixing of nuclear shapes.
\begin{table}[htb!]
\begin{center}
\begin{tabular}{ccccc}
\hline
\hline
  Nucleus   &  States        & Experiment &   $mp-mh$    & 5DCH    \\
\hline
 $^{28}$Si  &           &            &          &          \\
            & 2$^{+}_{1}$    &   1.779    &  1.993    &  2.469 \\
            & 4$^{+}_{1}$    &   4.618    &  5.372    &  6.446 \\
            & 0$^{+}_{2}$    &   4.980    &  4.409    & 10.591 \\
            & 3$^{+}_{1}$    &   6.276    &  6.365    &  \\
            & 0$^{+}_{3}$    &   6.690    &  8.760    &   \\
            & 4$^{+}_{2}$    &   6.888    &  7.769    &  \\
            & 2$^{+}_{2}$    &   7.381    &  7.280    &  7.395 \\
            & 2$^{+}_{3}$    &   7.416    &  8.353    &  \\
            & 3$^{+}_{2}$    &   7.799    &  8.124    &  \\
            & 2$^{+}_{4}$    &   7.933    &  8.551    & \\
            & 2$^{+}_{5}$    &   8.259    &  9.025    &  \\
            & 1$^{+}_{1}$    &   8.328    &  8.943    &  \\
            & 6$^{+}_{1}$    &   8.544    &  9.531    & 11.876 \\
            & 3$^{+}_{3}$    &   8.589    &  8.459    &  \\
            & 0$^{+}_{*}$    &   8.819    &  9.345    & \\
            & 5$^{+}_{1}$    &   8.945    &  9.424    &  \\
            & 0$^{+}_{*}$    &   8.953    &  9.845    &  \\
            & 4$^{+}_{3}$    &   9.164    & 10.120    &  \\
            & 3$^{+}_{4}$    &   9.315    &  9.894    &  \\
            & 2$^{+}_{6}$    &   9.381    &  9.883    &  \\
            & 4$^{+}_{4}$    &   9.417    & 10.719    &  \\
            & 2$^{+}_{7}$    &   9.479    &  9.952    &  \\
            & 1$^{+}_{2}$    &   9.496    &  9.756    &  \\
\hline
\hline
\end{tabular}
\end{center}
\caption{Same as Table \ref{tab5a} for $^{28}$Si; The symbol * means that spin-parity quantum numbers are not
assigned experimentally.}
\label{tab5b}
\end{table}

For the four isotopes, 3$^-$ is the lowest negative parity state. Its experimental energy is 6.789 MeV in $^{26}$Si,
6.879 MeV in $^{28}$Si, 5.487 MeV in $^{30}$Si and
5.288 MeV in $^{32}$Si. The decrease of its value gives a flavor of the increasing importance of negative parity
orbitals at low energy in $^{30}$Si and $^{32}$Si. However, for the description of positive parity states, negative
parity orbitals should play a role at energies higher than for the description of negative parity states as one has to introduce
at least $2p-2h$ excitations to produce a positive parity state.

Excitation energies calculated with the $mp-mh$ and 5DCH methods for $^{26-32}$Si are compared to experimental values in Tables \ref{tab5a}-\ref{tab6b} (the energies are expressed in MeV).
In Table \ref{tab6a}, a sixth column named $mp-mh_{s}$ has been added, showing $mp-mh$ results shifted by 2.5 MeV. Positive parity
has been assumed for the states in $^{26}$Si whose spins are not assigned experimentally, based on the
plausible hypothesis that the lowest negative parity state is the observed 3$^{-}$ one. The symbol ? means that
the state is not observed experimentally. For $^{28}$Si, the symbol * indicates that both spin and parity have not been
measured experimentally; again, making the assumption of positive parity, the spin is given by our model. For $^{32}$Si, because
of the energy of the lowest observed  3$^{-}$ state, positive parity is assumed for the fourth state at 5.220 MeV. Experimentally,
the spin is expected to be in the range from 1 to 4; our model predicts spin 3. Concerning the 4$^{+}_{*}$ state,
the experimental assignment is either 4$^{+}$ or 5$^{-}$.
\begin{table}[htb!]
\begin{center}
\begin{tabular}{cccccc}
\hline
\hline
  Nucleus   &  States        & Experiment &   $mp-mh$ & 5DCH  &  $mp-mh_{s}$     \\
\hline

 $^{30}$Si  &                 &                &               &     &      \\
            & 2$^{+}_{1}$    &   2.235    &    4.609  &  2.222 &  2.109    \\
            & 2$^{+}_{2}$    &   3.498    &    5.704  &  4.729 &  3.208    \\
            & 1$^{+}_{1}$    &   3.769    &    6.338  &            &  3.838    \\
            & 0$^{+}_{2}$    &   3.788    &    7.732  &  7.610 &  5.238   \\
            & 2$^{+}_{3}$    &   4.810    &    8.230  & 11.832&  5.730    \\
            & 3$^{+}_{1}$    &   4.831    &    6.709  &  8.056 &  4.209    \\
            & 3$^{+}_{2}$    &   5.231    &    7.904  &            &  5.404   \\
            & 4$^{+}_{1}$    &   5.279    &    7.539  &  5.691 &  5.039   \\
            & 0$^{+}_{3}$    &   5.372    &    8.950  &  8.585 &  6.450   \\
            & 2$^{+}_{4}$    &   5.614    &    9.262  &            &  6.762 \\
            & 4$^{+}_{2}$    &   5.950    &    8.911  &  8.714 &  6.441  \\
            & 2$^{+}_{5}$    &   6.538    &   10.186 &            &  7.686   \\
            & 0$^{+}_{4}$    &   6.642    &    9.030  &            &  6.530  \\
            & 3$^{+}_{3}$    &   6.865    &    8.831  &            &  6.331  \\
            & 2$^{+}_{6}$    &   6.914    &   10.594 &            &  8.094  \\
            & 5$^{+}_{1}$    &   6.998    &   10.926 &            &  8.426  \\
\hline
\hline
\end{tabular}
\end{center}
\caption{Same as Table \ref{tab5a} for $^{30}$Si.}
\label{tab6a}
\end{table}

Theoretical results are provided for $\sim\,$20 states in each isotope, except for $^{32}$Si where the spin and parity of some
states with excitation energies larger than $\sim\,$5.5 MeV have not been firmly assigned experimentally.
All excited configurations from the $sd$-shell have been introduced in the $mp-mh$ wave functions,
up to $6p-6h$ on the proton side and $2p-2h,4p-4h,6p-6h$ on the neutron side depending on the isotopes.
Then, the presented results contained up to $10p-10h$ configurations in $^{26}$Si and $^{30}$Si,
$12p-12h$ configurations in $^{28}$Si, and $8p-8h$ configurations in $^{32}$Si.

In the case of $^{26}$Si, one sees that there is a very good agreement between experimental and $mp-mh$ configuration mixing energies,
whatever the spin and the excitation energy. Concerning the 5DCH approach, the energy of the 2$^+_1$ state is found too high by
$\sim\,$700 keV and the energies of the other excited states are also strongly overestimated (by several MeV).
In the case of $^{28}$Si, despite the inversion between the 4$^+_1$ and 0$^+_2$ levels, quite good agreement with experiment is obtained by the $mp-mh$ approach. Again, the 5DCH approach tends to overestimate excitation energies. However, we note that both
theoretical approaches describe the 2$^+_1$ state within the same accuracy as in $^{26}$Si and that the energy of the 2$^+_2$ state is particularly well reproduced.

In $^{30}$Si, as seen from Table \ref{tab6a}, the energy of the 2$_{1}^{+}$ is overestimated by $\sim\,$2.5 MeV with the $mp-mh$ configuration mixing.
What is even more surprising is that all states appear to be shifted upwards.
Actually, reduction of all the excitation energies by $\sim$2.5 MeV gives a much better agreement
with experiment. This can be seen from the last column ``$mp-mh_{s}$" of Table \ref{tab6a} that gives the
values provided by
the $mp-mh$ configuration mixing approach minus 2.5 MeV. Then, the discrepancy with experiment is reduced to $\sim\,$0.69 MeV (on average), and the theoretical level sequence becomes very similar to the experimental one. We have checked that
this global shift cannot be removed by adding the $1f_{7/2}$ shell to the valence space.
Its origin will be discussed later in this subsection.

On the other hand, the energies of several states in this nucleus are well reproduced by the 5DCH approach, in particular those
of the 2$_{1}^{+}$ and 4$_{1}^{+}$ states.
Let us note in this respect that the 5DCH method does not explicitly make use of the matrix elements of the residual interaction
between excited configurations, exploiting instead the quadrupole
deformation properties of the mean field. As seen and discussed in relation to Fig. \ref{fig1b}, following the chemical potential, the deformation is able to catch part of the correlation information
coming from upper spherical shells directly or through the mixing of deformed orbitals (see discussion in
section \ref{section2A} on the evolution of single-particle orbitals).
In the case of $^{32}$Si, both theoretical methods provide a good description of the selected experimental states
(except for the 0$^+_2$ with the $mp-mh$ configuration mixing method and the 2$^+_3$ with the 5DCH approach).

The results obtained with the $mp-mh$ configuration mixing indicate that this approach is capable of reproducing quite well the low-energy spectroscopy of $^{26}$Si and $^{28}$Si
and to a lesser extent, the one of $^{32}$Si.
\begin{table}[htb!]
\begin{center}
\begin{tabular}{ccccc}
\hline
\hline
  Nucleus   &  States        & Experiment &   $mp-mh$    & 5DCH    \\
\hline
 $^{32}$Si  &           &            &          &          \\
            & 2$^{+}_{1}$    &   1.941    &    1.959       &  2.215 \\
            & 2$^{+}_{2}$    &   4.230    &    4.871       &  5.014 \\
            & 0$^{+}_{2}$    &   4.984    &    6.810       &  5.318 \\
            & 3$^{+}_{*}$    &   5.220    &    6.004       &            \\
            & 2$^{+}_{3}$    &   5.412    &    6.758       &  9.335 \\
            & 4$^{+}_{*}$    &   5.502    &    6.567       &  5.470 \\
\hline
\hline
\end{tabular}
\end{center}
\caption{Same as Table \ref{tab5a} for $^{32}$Si. The symbol * indicates that spin and/or parity is/are not assigned experimentally.}
\label{tab6b}
\end{table}
The mean deviations between theory ($mp-mh$) energies and experimental ones are found to be $\sim\,$369 keV in $^{26}$Si,
$\sim\,$653 keV in $^{28}$Si and $\sim\,$946 keV in $^{32}$Si. The increase of the deviation between $^{26}$Si and $^{28}$Si
can be attributed to the systematic overestimation obtained in the calculation of $^{28}$Si highest levels.
The same phenomenon holds for $^{32}$Si.
As pointed out previously, this effect can be partly ascribed to the absence of the $1f_{7/2}$ shell,
and, to a smaller extent, of the $2p_{3/2}$ shell in the present calculations.
Nonetheless, the agreement with experiment of $mp-mh$ energies can be considered as rather encouraging,
considering the fact that the D1S Gogny interaction has not been devised to describe the kind of general correlations introduced
in a multiconfiguration approach. In particular,
the proton-neutron matrix elements between excited configurations given by this interaction have not been constrained.

The following discussion is dedicated to the understanding of the origin of the
shift obtained with the $mp-mh$ configuration mixing method for excited states in $^{30}$Si.
In a schematic way, both energy gaps between single-particle orbitals and coupling matrix elements (ME) between configurations are the key quantities that drive the low-lying spectroscopy. One can infer that a downward shift can be obtained either by decreasing gaps (an effect similar to the monopole shifts pointed out in the shell model approach [24-26]), and/or by varying coupling matrix elements. Proton-neutron matrix elements are suspected to be
mainly responsible for the energy shift encountered in $^{30}$Si.
In fact, by changing ``by hand" the values of selected proton-neutron ME implying the spin-orbit partners 1$d_{3/2}$ and 1$d_{5/2}$ and using them in realistic $mp-mh$ configuration mixing calculations (which produces a modification of the energy of the 1$d_{3/2}$ shell), one can derive excited states in $^{30}$Si which are 
in reasonably good agreement with experiment, with deviations similar to the ones found in $^{28}$Si. It 
is important to note that such changes in  matrix elements essentially effect only the $^{30}$Si spectrum. 
In particular no significant modification of  $^{28}$Si spectrum is observed.
In addition, as we will see in section \ref{section3} where the chaotic behaviour of highly excited Slater determinants is studied, too strong couplings are found essentially in $K=0$ cases,
where common proton-neutron matrix elements are involved.

At this stage of our analysis, one has to recall that the
general formalism of the $mp-mh$ configuration mixing method exposed in section
\ref{section1} implies that not only the secular equation (\ref{eq4}) has to
be solved but also Eq. (\ref{eqad1}). These two equations in principle provide the ``best"
single-particle representation, i.e. the one that
minimizes the total energy consistent with correlations.
Clearly, the solution of Eq. (\ref{eqad1}) may introduce
modifications on both single-particle energies and coupling ME.
However, discussing the kind of renormalization produced by
Eq. (\ref{eqad1}) is far beyond the scope of the present paper, and it will be
left to a dedicated study.
Let us simply mention here that, in the context of the present
work, introducing a tensor term in the effective interaction we would probably reproduce
the right energy evolution of spin-orbit partners \cite{marta}.
In addition, a crude comparison between $sd$-shell ME calculated from the D1S Gogny
interaction and from the USD interaction used in the shell model \cite{wildenthal} displays
large discrepancies essentially in the $T=0$ channel, where
the renormalization effects are expected to be
the largest. The average difference is equal to $\sim$0.3 MeV in $T=1$ channel and
$\sim$1.5 MeV in $T=0$ one. The large difference found in $T=0$ channel is attributed essentially
to two ME and comes from the lack of tensor term by comparing the different contributions to
ME (central, spin-orbit and tensor).

\begin{figure}[htb!]
\begin{center}
\includegraphics[height=7cm, angle=0]{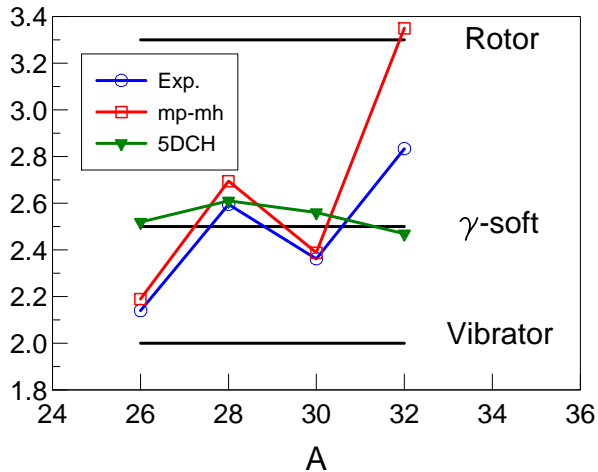}
\end{center}
\caption{(Color online) Experimental and calculated ($mp-mh$ and 5DCH) $E(4^{+}_{1})/E(2^{+}_{1})$ ratios. For $^{30}$Si, the shifted for $mp-mh$ configuration
mixing energies have been used (see text).}
\label{fig20d}
\end{figure}
\begin{table}[htb!]
\begin{center}
\begin{tabular}{cccc}
\hline
\hline
  Nucleus & $\frac{E(4^{+}_{1})}{E(2^{+}_{1})}_{{\rm exp.}}$ & $\frac{E(4^{+}_{1})}{E(2^{+}_{1})}_{mp-mh}$ & $\frac{E(4^{+}_{1})}{E(2^{+}_{1})}_{{\rm 5DCH}}$    \\
\hline
 $^{26}$Si  &    2.140                                        & 2.190                                              &  2.520  \\
 $^{28}$Si  &    2.595                                        & 2.695                                              &  2.611 \\
 $^{30}$Si  &    2.362                                        & 2.389$^s$                                      &  2.561  \\
 $^{32}$Si  &    2.834                                        & 3.350                                              &  2.469            \\
\hline
\hline
\end{tabular}
\end{center}
\caption{ Numerical values of experimental and calculated ($mp-mh$ and 5DCH) $E(4^{+}_{1})/E(2^{+}_{1})$ ratios. The superscript
``s" for $mp-mh$ ratios in $^{30}$Si means that the energies shifted by 2.5 MeV have been used, see text.}
\label{tab6c}
\end{table}

In conclusion of this part, one can say that, in the four silicon isotopes, most states contain more complex
correlations than the usual collective quadrupole/rotational ones. In order to identify the nature of the ground state band,
we have calculated the energy ratio ($4^+_1 / 2^+_1$) whose value is a standard indicator of the vibrational, rotational or $\gamma$-soft nature of nuclei.
Results for experiment, $mp-mh$ configuration mixing and 5DCH methods are displayed in Fig. \ref{fig20d}. Numerical values are reported in Table \ref{tab6c} for experiment, $mp-mh$ configuration mixing, and 5DCH methods. From experiment, one
observes that $^{26}$Si is close to the vibrational limit
($(4^+_1 / 2^+_1)_{{\rm vib}}=2$), while $^{28}$Si and $^{30}$Si are more $\gamma$-soft nuclei
($(4^+_1 / 2^+_1)_{\gamma}=2.5$)). In $^{32}$Si, this ratio
increases towards the rotor limit ($(4^+_1 / 2^+_1)_{{\rm rot}}=3.3$)).
The results obtained with the $mp-mh$ configuration
mixing show that data in $^{26}$Si and $^{28}$Si are well reproduced and display the experimental trend.
For $^{30}$Si, the shifted values for the $mp-mh$ configuration mixing approach have been used.
The significant overestimation in $^{32}$Si, where the 2$^{+}_{1}$ energy is well reproduced, comes from the fact that the 4$^{+}_{1}$ energy
is slightly overestimated.

\subsection{Role of the proton-neutron residual interaction}\label{pn}

This part discusses the role of the proton-neutron residual interaction and the importance of using an effective nucleon-nucleon interaction that manifests
good properties in the $T=0$ channel when using $mp-mh$ configuration mixing methods. As a benchmark nucleus, we have chosen $^{28}$Si.
Table \ref{tab1} lists the first seven excited states of $^{28}$Si. The experimental values of excitation energies
and the theoretical ones ($mp-mh$) are given in columns 2 and 3. Columns 4 and 5 display the results obtained within
the $mp-mh$ configuration mixing approach when the residual proton-neutron interaction is turned off.
As the Hamiltonian in use is not exactly isospin-invariant (a Coulomb term is included),
this symmetry breaking leads to a small difference (less than $\sim\,$300 keV) for proton and neutron solutions,
noted $mp-mh_{\pi}$ and $mp-mh_{\nu}$, respectively.
\begin{table}[h!]
\begin{center}
\begin{tabular}{ccccc}
\hline
\hline
       States  & Experiment &       $mp-mh$   &  $mp-mh_{\pi}$ & $mp-mh_{\nu}$  \\
\hline
       2$^{+}_{1}$    &   1.779    &  1.993    &  5.733               & 5.831 \\
       4$^{+}_{1}$    &   4.618    &  5.372    &  6.553               & 6.712 \\
       0$^{+}_{2}$    &   4.980    &  4.409    &  9.588              & 9.651 \\
       3$^{+}_{1}$    &   6.276    &  6.365    &  9.732              &  ? \\
       0$^{+}_{3}$    &   6.690    &  8.759    &  9.873              & 9.893  \\
       4$^{+}_{2}$    &   6.888    &  7.769    &    ?                    &   ?  \\
       2$^{+}_{2}$    &   7.380    &  7.280    & 10.283              & 10.415 \\
\hline
\hline
\end{tabular}
\end{center}
\caption{Excitation energies (in MeV) of low-lying states in $^{28}$Si calculated
with the variational $mp-mh$ configuration mixing method with and without residual proton-neutron interaction (see text).}
\label{tab1}
\end{table}

The sensitivity of the excitation energies to the proton-neutron residual interaction
depends on the nature of states. For example, this interaction
brings the energy of the 2$^{+}_{1}$ state from $\sim\,$6 MeV to $\sim\,$2 MeV;
its importance for the structure of correlated wave functions
is illustrated in Table \ref{tab2} where the components of the wave functions
of the 0$^{+}_{1}$ and 0$^{+}_{2}$ states are listed in two cases. We define the quantity $W_{n}$ that measures
the correlation content of the wave functions in terms of the order of excitation $n$, namely $0p-0h,\, 1p-1h, 2p-2h, ...$
For a given eigenfunction $\vert \Psi_{\beta} \rangle$,
\begin{equation}
\dspt W_{n}^{\beta}= \sum_{k_{n}} |A_{k_{n}}^{\beta}|^{2},
\label{eqqq1}
\end{equation}
where $k_{n}$ represents the Slater determinant components with $np-nh$ excitations ($n=n_{\pi}+n_{\nu}$).
\begin{table}[h!]
\begin{center}
\begin{tabular}{cccccccccc}
\hline
\hline
 State & Case & $W_{0}$ & $W_{1}$ &$W_{2}$ & $W_{3}$  &$W_{4}$ & $W_{5}$ & $W_{6}$ & $W_{7}$\\
\hline
 0$^{+}_{1}$ & (a) & 34.48 & 0.00 & 34.77 & 11.55 & 12.78 & 4.25 & 1.76 & 0.34 \\
                    & (b) & 93.77 & 0.00 & 6.03 & 0.02 & 0.17 & 0.00 & 0.00 & 0.00  \\
 0$^{+}_{2}$ & (a) & 43.70 & 0.00 & 12.26 & 14.12 & 16.37 & 8.68 & 3.82 & 0.90   \\
                    & (b$_{\pi}$) &  0.00 & 91.51& 2.82 & 5.34 & 0.20 & 0.12 & 0.00 & 0.00  \\
                    & (b$_{\nu}$) &  0.00 & 91.39& 3.25 & 5.04 & 0.21 & 0.11 & 0.00 & 0.00  \\
\hline
\hline
\end{tabular}
\end{center}
\caption{Weights $W_{n}$ (n$\le$8) calculated for 0$^{+}_{1}$ and 0$^{+}_{2}$ states in $^{28}$Si.
Case (a) corresponds to the full $mp-mh$ calculation and case (b) to a calculation without residual
proton-neutron interaction. The index $\pi$ ($\nu$) specifies the proton (neutron) solution.}
\label{tab2}
\end{table}
Case (a) in Table \ref{tab2} corresponds to a full calculation and case (b) to a calculation without residual proton-neutron interaction.
The wave functions 0$^{+}_{1}$ and  0$^{+}_{2}$ have quite different structures.
For the ground state, excited $2p-2h$ configurations play a role as important as the initial $0p-0h$ configuration.
When the proton-neutron residual interaction is turned off, most of the correlations disappear.
One sees that the $2p-2h$ configurations built from $1p-1h$ proton excitations combined with $1p-1h$
neutron excitations are essential for the description of the ground state. For the excited state, the absence of proton-neutron
interaction is even worse as it destroys fully the $0p-0h$ component and produces the solutions based on a $1p-1h$
proton or on a $1p-1h$ neutron configuration.
One observes that the proton-neutron interaction brings a lot of fragmentation in the wave functions, hence collectivity.
The precise knowledge of this residual interaction is therefore mandatory in $mp-mh$ configuration mixing approaches, in particular when calculations are performed at sphericity.
Consistently with the results for the wave functions, occupation probabilities display strong changes as can be seen
in Table \ref{tab3} in the case of proton and neutron $1d_{5/2}$, $2s_{1/2}$ and $1d_{3/2}$ orbitals.

\begin{table}[h!]
\begin{center}
\begin{tabular}{cccccccc}
\hline
\hline
 State & Case & $\pi d_{5/2}$ & $\nu d_{5/2}$ & $\pi s_{1/2}$ & $\nu s_{1/2}$ & $\pi d_{3/2}$ & $\nu d_{3/2}$ \\
\hline
 0$^{+}_{1}$ & (a) & 5.053 & 5.047 & 0.603 & 0.606 & 0.344 & 0.347 \\
                    & (b) & 5.939 & 5.932 & 0.018 & 0.020 & 0.043 & 0.047  \\
 0$^{+}_{2}$ & (a) & 4.960 & 4.978 & 0.716 & 0.698 & 0.324 & 0.324 \\
                    & (b$_{\pi}$) & 4.941 & 5.912 & 1.007 & 0.037 & 0.050 & 0.050  \\
                    & (b$_{\nu}$) & 5.935 & 4.935 & 0.033 & 1.008 & 0.046 & 0.057  \\
\hline
\hline
\end{tabular}
\end{center}
\caption{Proton and neutron occupation probabilities of the $d_{5/2}$, $s_{1/2}$ and $d_{3/2}$ orbitals, for
0$^{+}_{1}$ and 0$^{+}_{2}$ states in $^{28}$Si. Case (a) corresponds to a full calculation and case (b) to a
calculation without residual proton-neutron interaction.}
\label{tab3}
\end{table}

\section{Statistical properties of highly excited configurations}\label{section3}

One of the main issues raised by multiconfiguration approaches is the number of relevant configurations
for describing low-lying states. Because of the proton-neutron
excitations, this number can rapidly explode. In realistic calculations, one has to think about
truncations based on physical arguments. When going beyond the mean-field with the $mp-mh$ approach, it is assumed
that short-range correlations have already been taken into account through the effectiveness of the nucleon-nucleon
interaction used (the D1S Gogny interaction in our case).
Our aim is to treat explicitly the long-range correlations corresponding to the attractive part
of the nucleon-nucleon interaction. Two standard
types of truncation can be proposed, independently of the choice (or not) of a valence space: a truncation on the order of the
excitations and/or a truncation on the configuration excitation energies.
Both types of truncations seem to be reasonable in the present context. In order to define appropriate truncations based on relevant physics
argument, we discuss below the behavior of highly excited configurations.

We have followed the direction used in Refs. \cite{ec1,ec2,ec3,ec4,ec5} where the analysis was based on the properties
of the strength function associated with the Slater determinants.
Using second quantization and the standard Wick's theorem decomposition, the Hamiltonian $\hat{\cal{H}}$,
Eq. (\ref{eq5}), is the sum of an independent particle part $\hat{\cal{H}}_{0}$ (one-body) and a residual part
$\hat{\cal{H}}'$ (one-body and two-body),
\begin{equation}
\hat{{\cal{H}}}= \hat{{\cal{H}}}_{0} + \hat{{\cal{H}}}'.
\label{eq6}
\end{equation}
The eigenfunctions $|k\rangle$ of the unperturbed Hamiltonian ${\cal{H}}_{0}$,
\begin{equation}
{\cal{H}}_{0} \vert k \rangle = \epsilon_{k} \vert k \rangle,
\label{eq8}
\end{equation}
describe noninteracting fermions.
In the basis $\vert k \rangle$, the residual interaction $\hat{{\cal{H}}}'$ has both diagonal,
$\overline{\hat{{\cal{H}}}}$, and off-diagonal, $\widetilde{\hat{{\cal{H}}}}$, matrix elements:
${\hat{\cal{H}}}'= \overline{\hat{{\cal{H}}}} + \widetilde{\hat{{\cal{H}}}}$. Full diagonalization
leads to the stationary states $\vert \alpha \rangle$ and their energies ${\cal E}_{\alpha}$,
\begin{equation}
{\hat{\cal{H}}} \vert \alpha \rangle = {\cal{E}}_{\alpha} \vert \alpha \rangle.
\label{eq9}
\end{equation}
The eigenfunctions $|\alpha\rangle$ may have a complicated structure in the original basis $|k\rangle$,
\begin{equation}
\vert \alpha \rangle = \sum_{k} A^{\alpha}_{k} \vert k \rangle.
\label{eq7}
\end{equation}
where $A$ is a unitary matrix:
\begin{equation}
\sum_{k} A^{\alpha}_{k} A^{\alpha'}_{k} = \delta^{\alpha \alpha'},~~~~~~
\sum_{\alpha} A^{\alpha}_{k} A^{\alpha}_{k'} = \delta_{k k'}.
\label{eq10}
\end{equation}
A completely delocalized wave function $\vert \alpha \rangle$ would have a number of relevant components close to the total dimension $N$ of the multiconfiguration space (for given exact quantum numbers). In this limit the typical magnitude of each component is $1/\sqrt{N}$.
In general, a number $N_{\alpha}$ of principal components
$\vert k \rangle$ characterizes the delocalization of a state $\vert \alpha \rangle$ in the given basis (\ref{eq8}).
Indeed, a two-body interaction can not couple configurations differing by more than two particle states which implies a
band-like Hamiltonian matrix, favoring the localization of eigenfunctions in the Hilbert space.
Conversely, the fragmentation of simple basis states over the energy spectrum can be provided by the strength function
defined by the quantity
\begin{equation}
F_{k}(E)= \langle k \vert \delta(E- \hat{H}(\rho) ) \vert k \rangle =
\sum_{\alpha} |A^{\alpha}_{k}|^{2} \delta(E-{\cal E}_{\alpha}).
\label{st1}
\end{equation}
\begin{figure}[htb!]
\begin{center}
\includegraphics[height=7.0cm, angle=0]{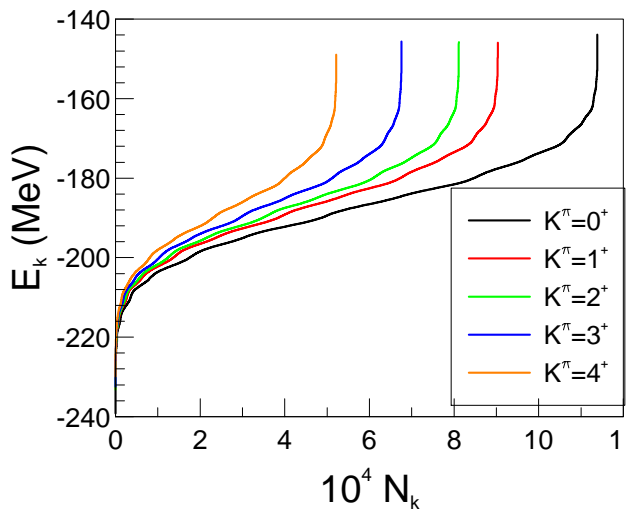}
\end{center}
\caption{(Color online) Centroid energies (in MeV) for $^{28}$Si. The x-axis corresponds to the number of centroids ordered by increasing energies.}
\label{fig10}
\end{figure}
The strength function contains rich information but requires the knowledge of the full nuclear spectrum. Fortunately,
one can study the main characteristics of the system from the first and the second moment
of the strength function which does not require the actual diagonalization of ${\cal{H}}$.
These two moments are the centroid, $E_{k}$, and dispersion,
$\sigma_{k}$, of the state distribution, given by
\begin{equation}
E_{k}= \int dE~EF_{k}(E)= {\cal{H}}_{kk} = \epsilon_{k} + {\overline{\cal{H}}}_{kk},
\label{eq11}
\end{equation}
and
\begin{equation}
\sigma^{2}_{k} = \int dE~(E-\overline{E}_{k})^2 F_{k}(E)= \sum_{l(\ne k)} ({\cal{H}}'_{lk})^2,
\label{eq13}
\end{equation}
respectively.
The centroid $\overline{E}_{k}$ coincides with the unperturbed energy ${\cal{H}}_{kk}$, whereas the
dispersion depends only on the off-diagonal matrix elements ${\cal{H}}'_{lk}$. \\

\begin{figure*}[htb!]
\begin{center}
\includegraphics[height=6.0cm, angle=0]{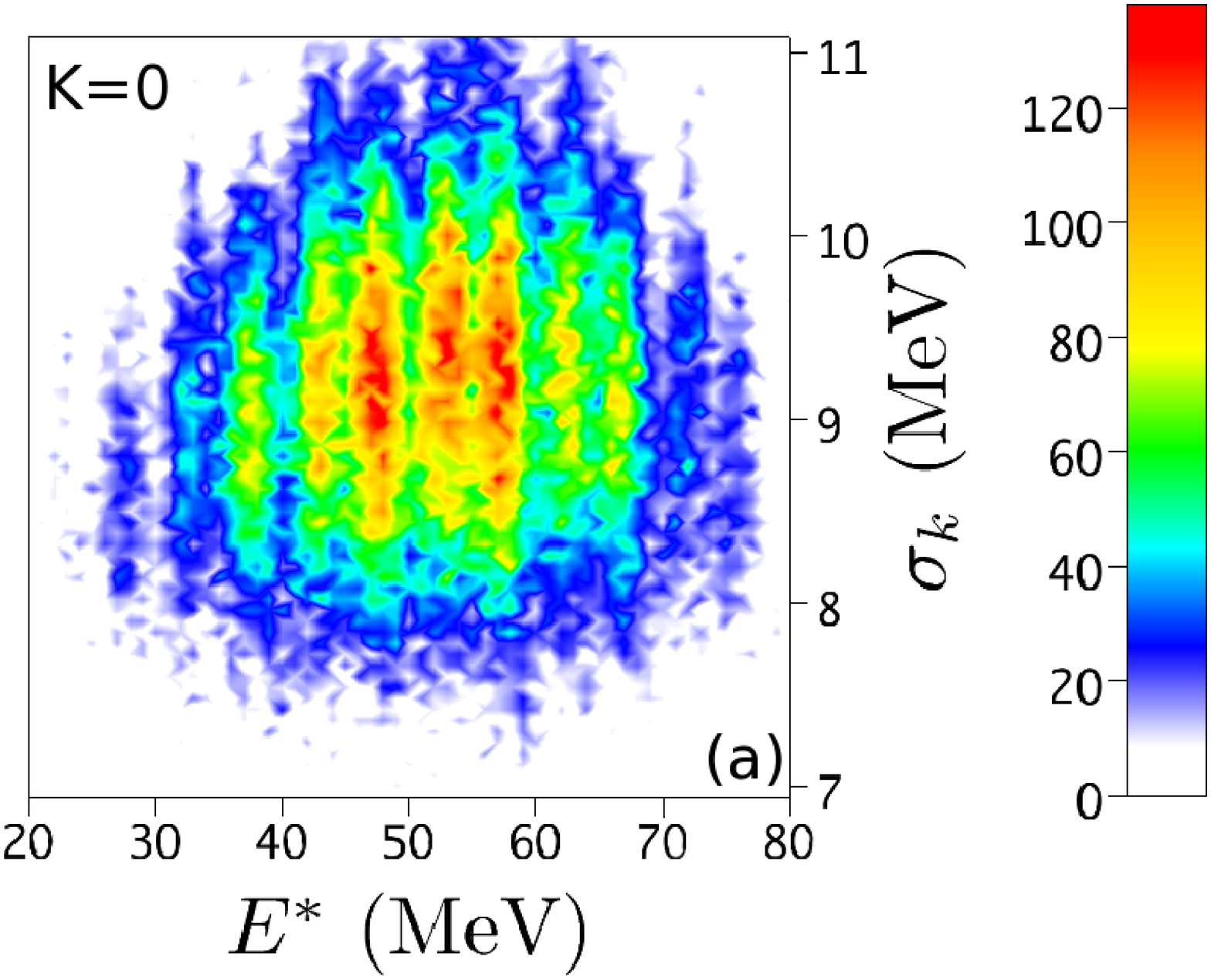} \hspace{0.5cm}
\includegraphics[height=6.0cm, angle=0]{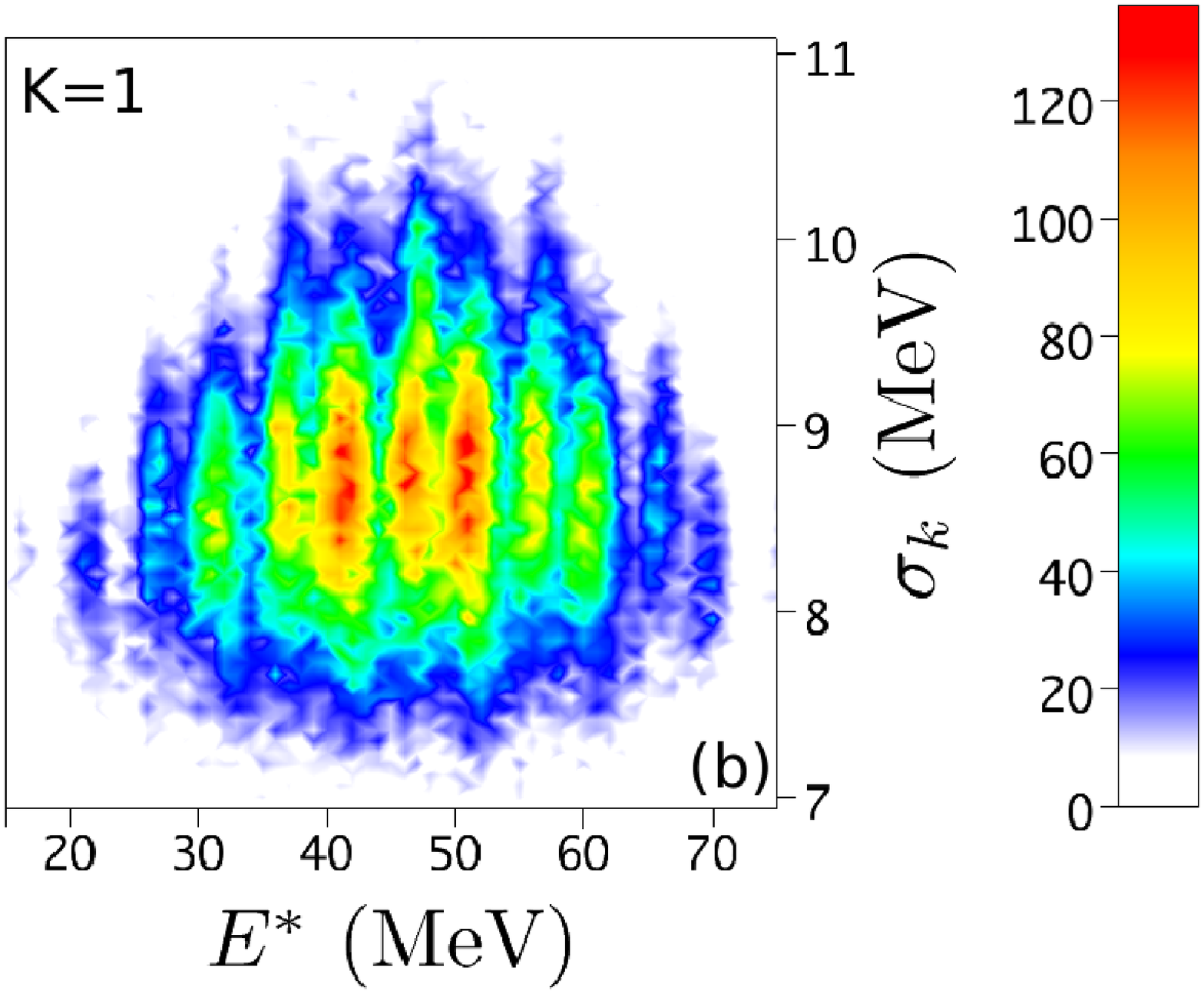}
\includegraphics[height=6.0cm, angle=0]{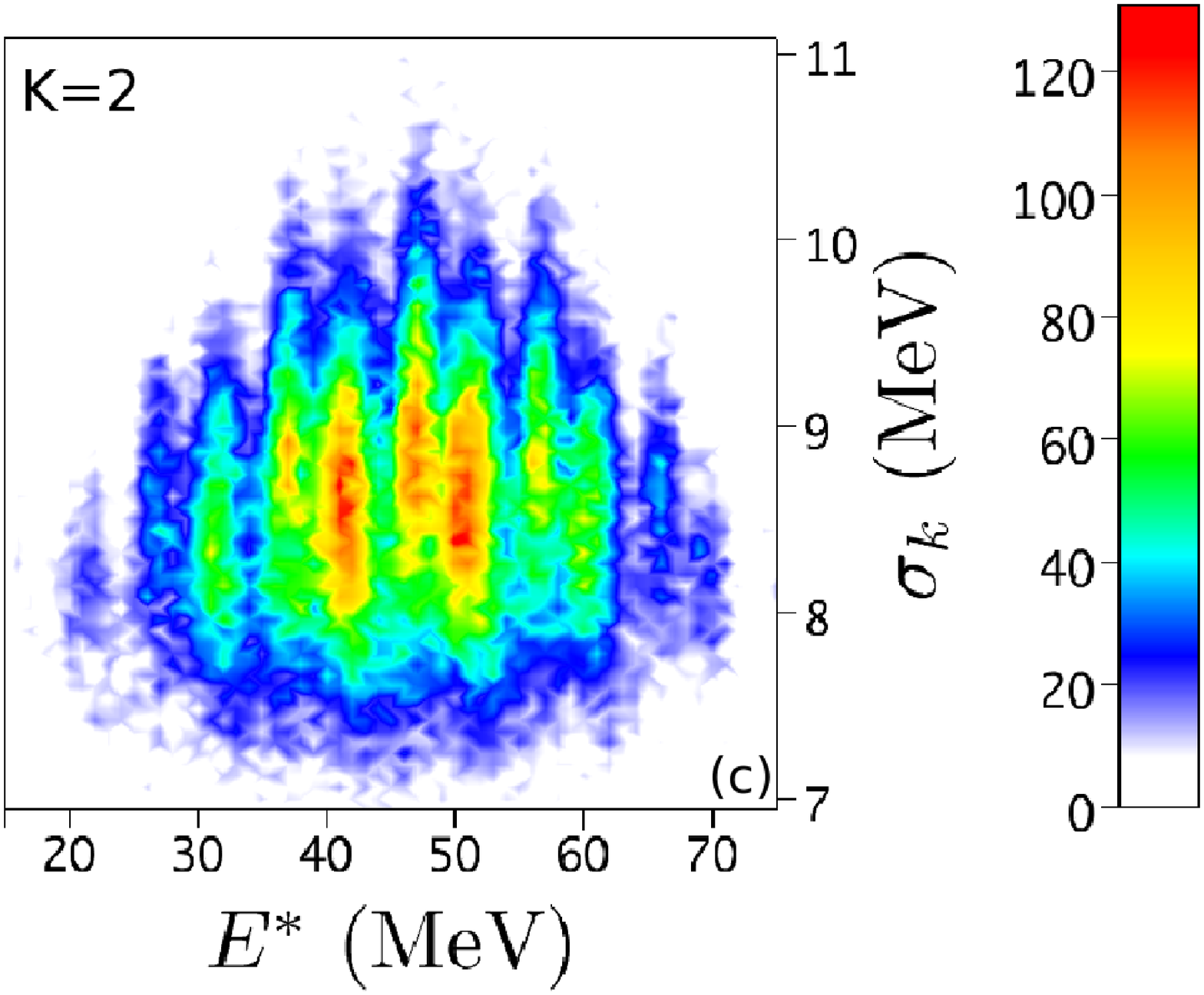} \hspace{0.5cm}
\includegraphics[height=6.0cm, angle=0]{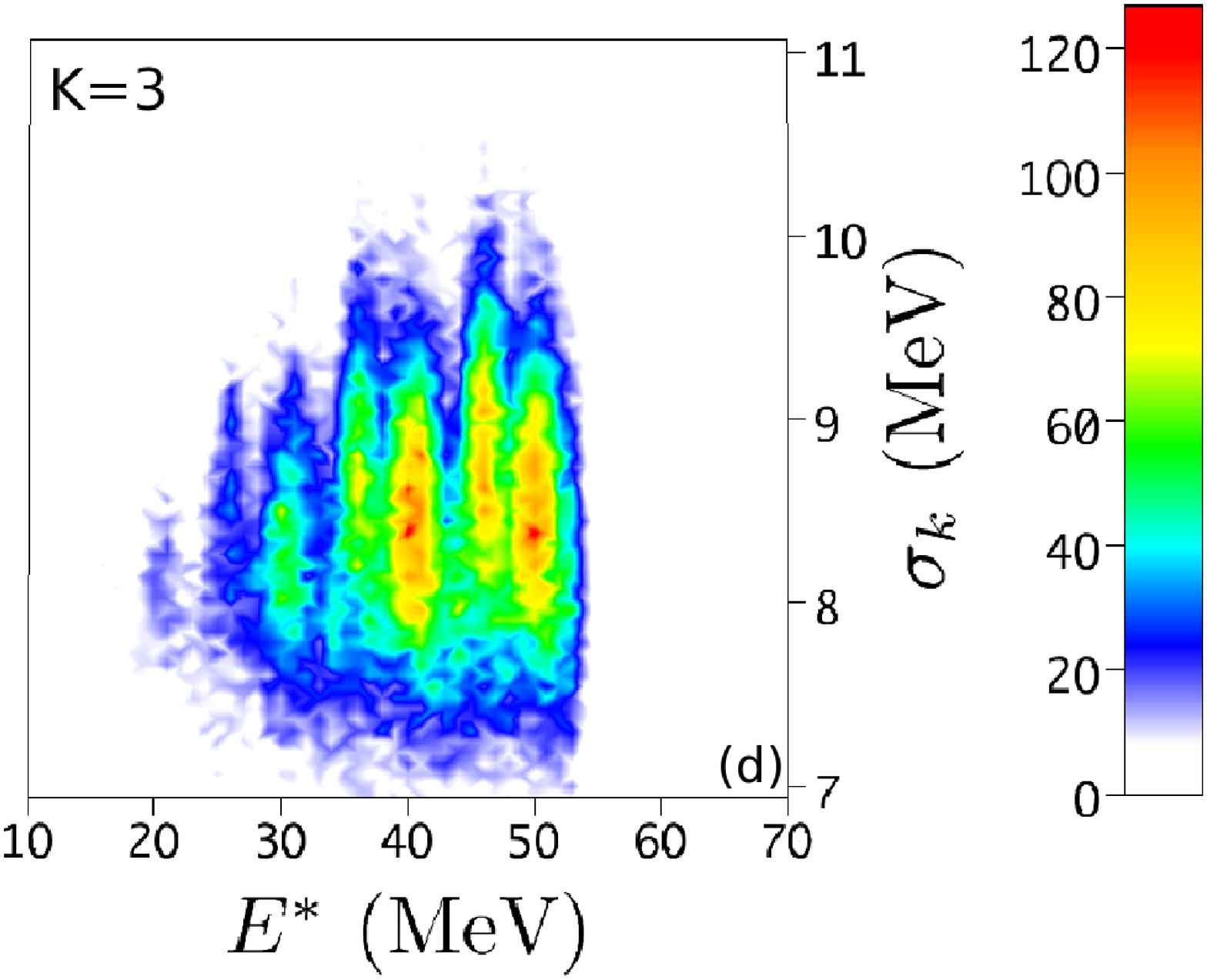}
\end{center}
\caption{(Color online) Dispersions $\sigma_k$ (in MeV) of $K^{\pi}=$ 0$^+$, 1$^+$, 2$^+$ and 3$^+$ Slater determinants for $^{28}$Si. The $x$-axis represents the excitation energies
$E^*$ of Slater determinants. The color code indicates the number of Slater determinants per bins of excitation energy and dispersion.}
\label{fig11}
\end{figure*}

Below we discuss the example of the $^{28}$Si isotope. Similar calculations have been done for other nuclei with similar conclusions. At this point, it is important to recall that the $mp-mh$ configuration mixing formalism presented in Section \ref{section1} has been developed, in practice, using an axially-deformed harmonic oscillator basis. In order to preserve the spherical symmetry, all the $mp-mh$ calculations displayed in this study have been done for $\beta=0$. In the following, our analysis is done in terms of different projections $K$ of a given
angular momentum $J$. All configurations of the $sd$-shell (up to $12p-12h$) have been introduced in the
wave functions; the maximum size of the Hamiltonian matrices that have been fully diagonalized was
$\sim\,$90 000 $\times$ 90 000.

Fig. \ref{fig10} displays the values of the centroid energies $E_{k}$,
where centroids are labeled by N$_{k}$ and ordered by increasing energy. One observes a characteristic behavior for
all values of $K$ from 0 to 4. The lowest centroid energy,
the one associated with the $0p-0h$ configuration, is -239.203 MeV. Only few configurations have small excitation energy. The level
density increases rapidly with excitation energy, and most configurations are located in the range $[-210;-160]$ MeV.
The final increase, beyond -160 MeV, is an artifact of the finite valence space that reduces the number of possible highly excited states.
\begin{figure}[h!]
\begin{center}
\includegraphics[height=7.0cm, angle=0]{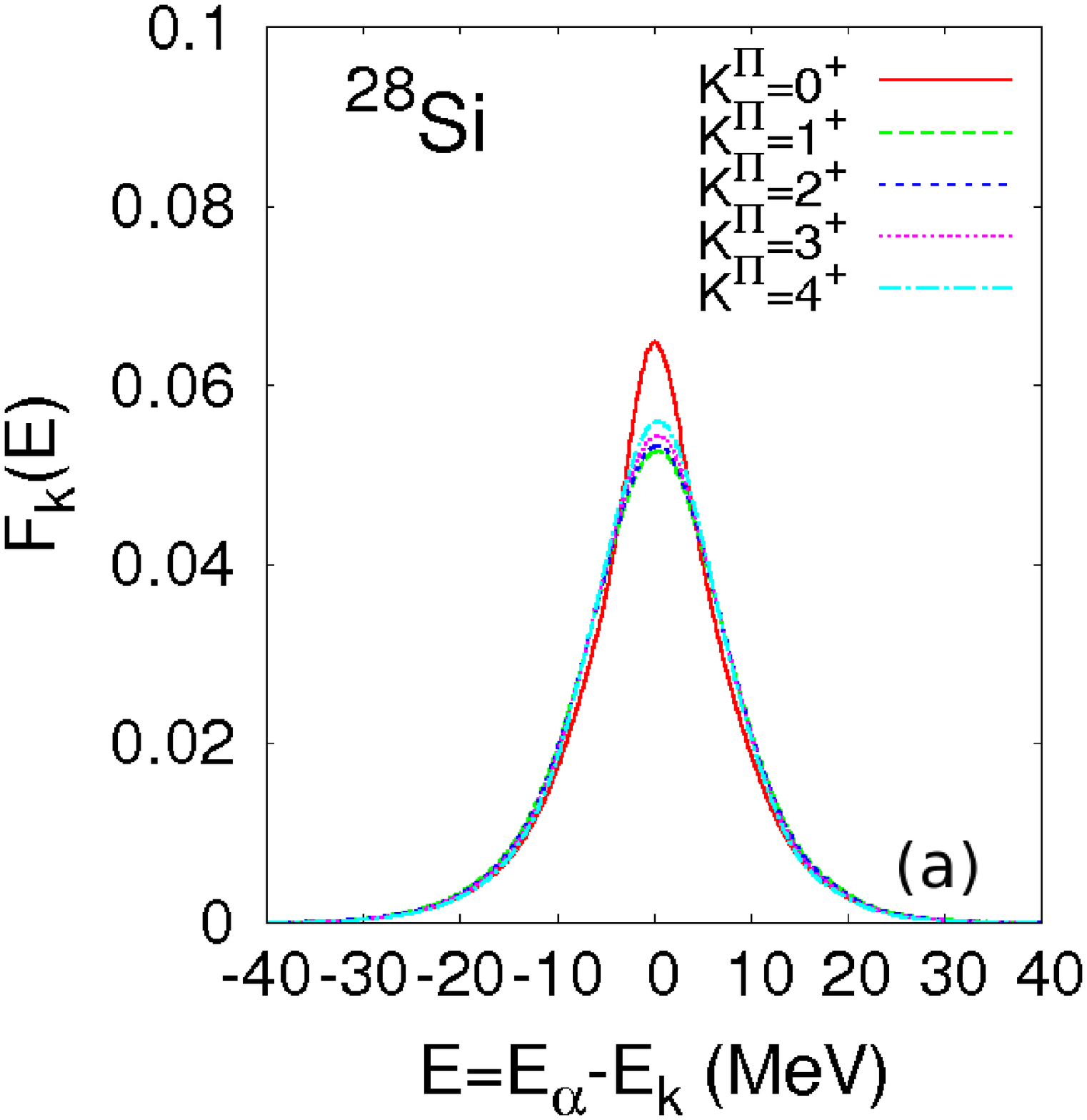}
\hspace{0.5cm} \includegraphics[height=7.0cm, angle=0]{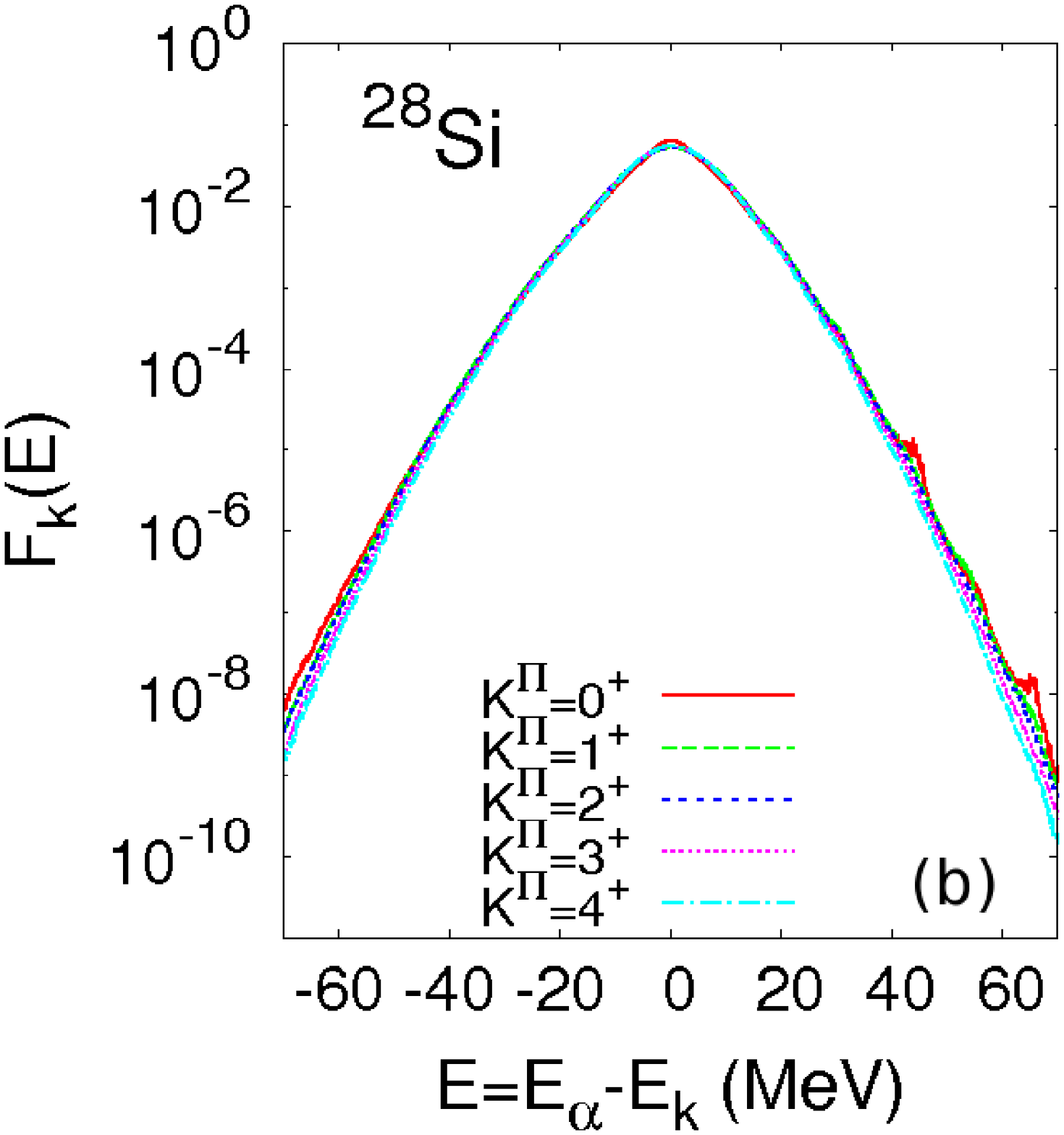}
\end{center}
\caption{(Color online) The $^{28}$Si strength functions associated with $K^{\pi}=0^+,\, 1^+,\, 2^+,\, 3^+$ and 4$^+$ components in linear scale (upper panel) and logarithmic scale (lower panel).
Logarithmic representation is shown in lower panel. Calculations have been performed with a 0.1 MeV bin.}
\label{fig7a}
\end{figure}

Fig. \ref{fig11} shows the dispersions $\sigma_{k}$ of Eq. (\ref{eq13}) for all configurations characterized
by $K= 0, 1, 2$ and 3 and their excitation energies $E^*$ (in MeV). The color code indicates the number of configurations in a given excitation energy and $\sigma_{k}$ bin. As expected, the structure of the
configuration distribution is different for different values of $K$ but most configurations are concentrated along a central line, and characteristic structures appear in the most dense areas. Such areas are more pronounced in the most dense zone in the $K=0$ case for all values of $E^*$.
A large majority of configurations display a dispersion $\sigma_{k}$ that is rather constant along the spectrum;
its mean value can be evaluated to be $\sim\,$9 MeV.
This important result which implies that strength functions can be characterized by essentially the same width
is the first indication of expected exponential convergence.
\begin{figure}[h!]
\begin{center}
\includegraphics[height=7.0cm, angle=0]{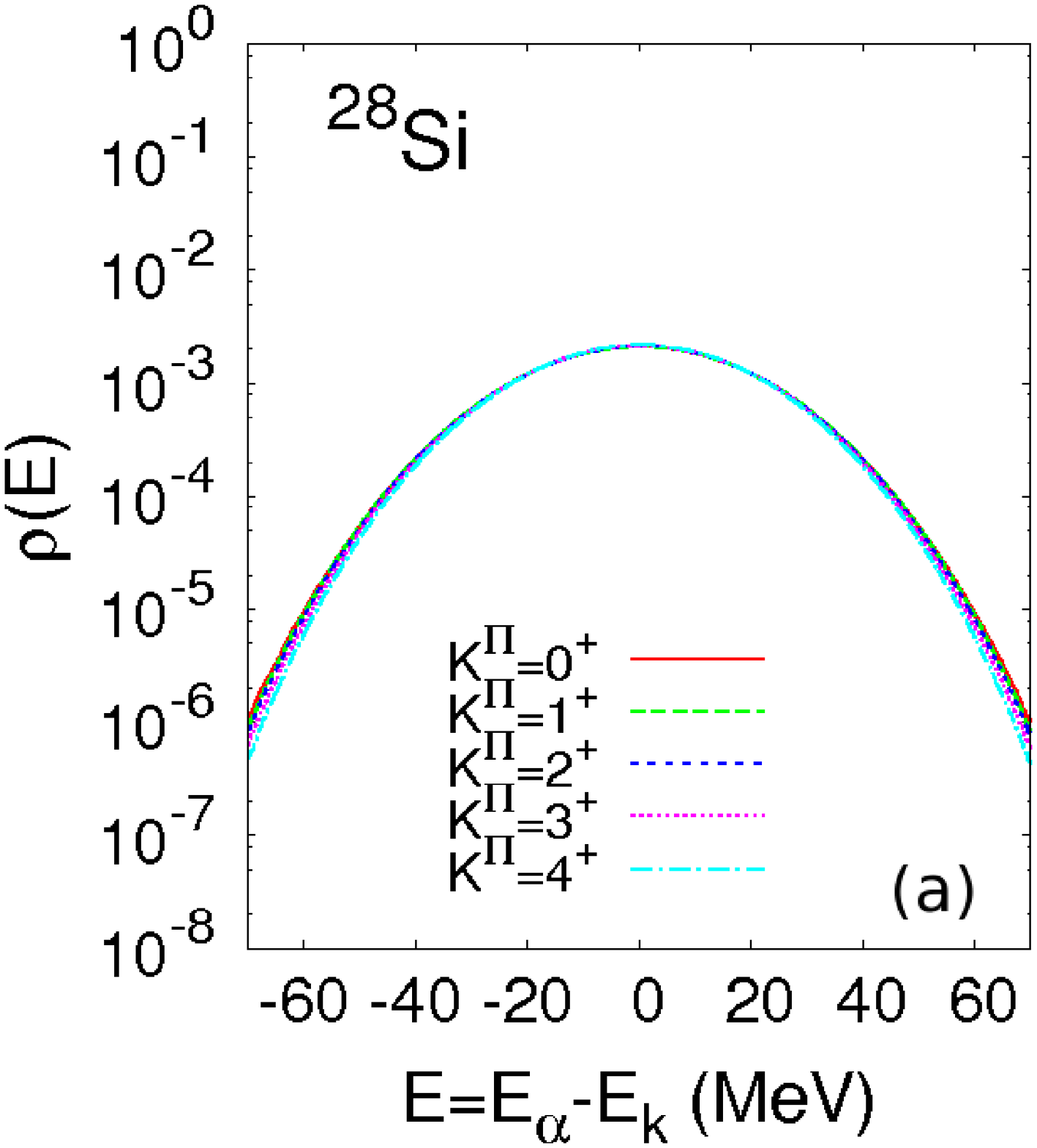}
\hspace{0.5cm} \includegraphics[height=7.0cm, angle=0]{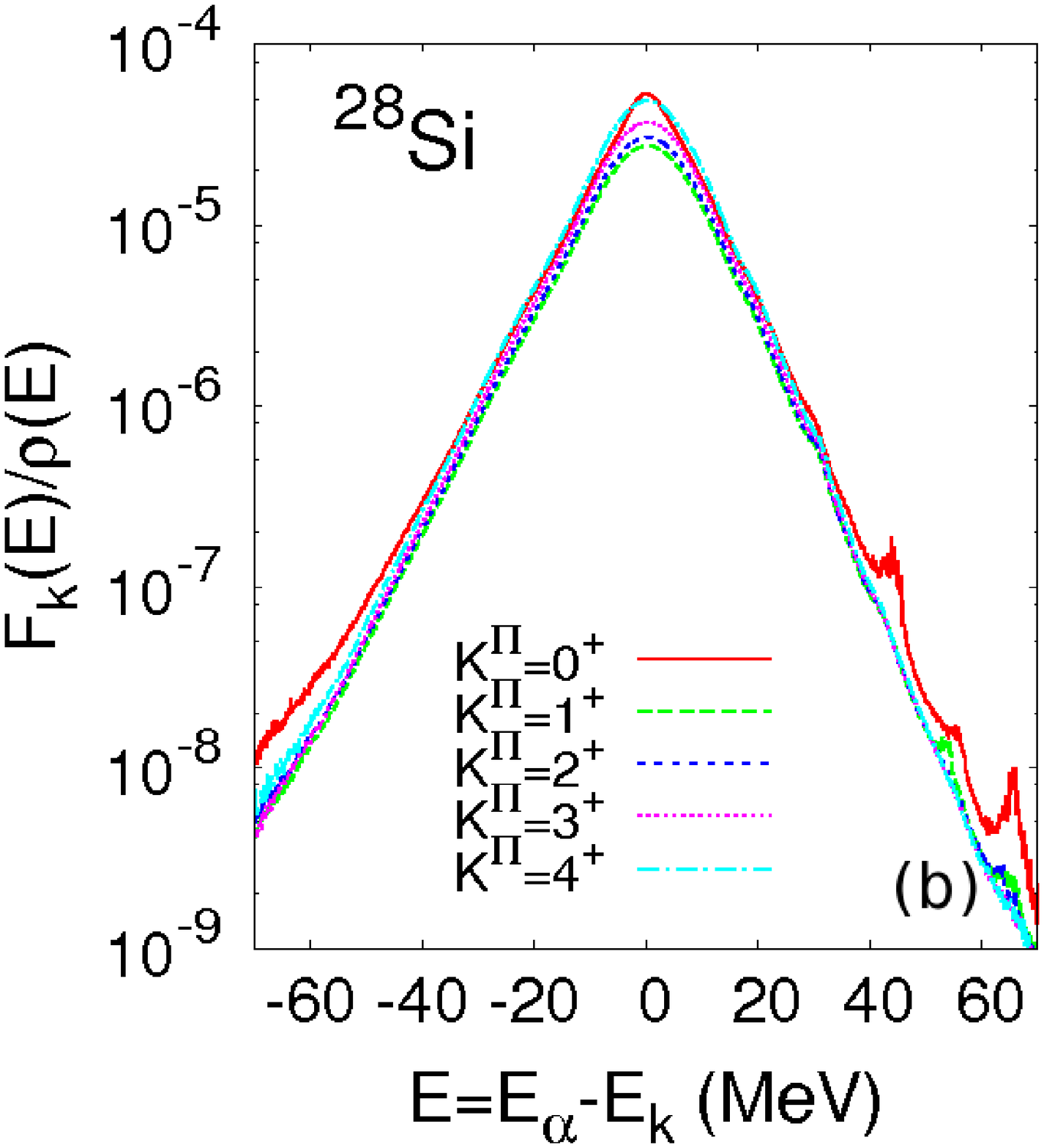}
\end{center}
\caption{(Color online) The $^{28}$Si Slater determinant distributions $\rho(E)$ (upper panel) and the $^{28}$Si locally normalized strength functions $\langle F_{k} (E) \rangle / \rho(E)$
(lower panel) for $K^{\pi}=0^+,\, 1^+,\, 2^+,\, 3^+$ and 4$^+$ components . Calculations have been performed with a 0.1 MeV bin.}
\label{fig7c}
\end{figure}

The strength function of $^{28}$Si is presented in Fig. \ref{fig7a} for different values of $K$ as a function
of the energy counted from the corresponding centroid, $E_{\alpha}-E_{k}$. The value of the bin is set to 0.1 MeV.
An average over all configurations has been done in order to reveal possible
generic behavior. The upper panel displays the strength function in linear scale and the lower panel in
logarithmic scale. The central part of the distribution is intermediate between a Gaussian and a Lorentzian one,
whereas the behavior in the wings is found to be a decreasing exponential. This is clearly visible
in the logarithmic scale. In agreement with \cite{ec1,ec2,ec3,ec4,ec5}, the coupling of highly excited configurations with
low-energy eigenfunctions therefore exhibits an exponential regime and the ``3 $\sigma$ "
rule characterizing the start of the exponential convergence, which is important in practical calculations
of the level density by methods of statistical spectroscopy \cite{senkov11}, seems to hold here also.
\begin{figure*}[htb!]
\begin{center}
\hspace{0.0cm}\includegraphics[height=6.5cm, angle=0]{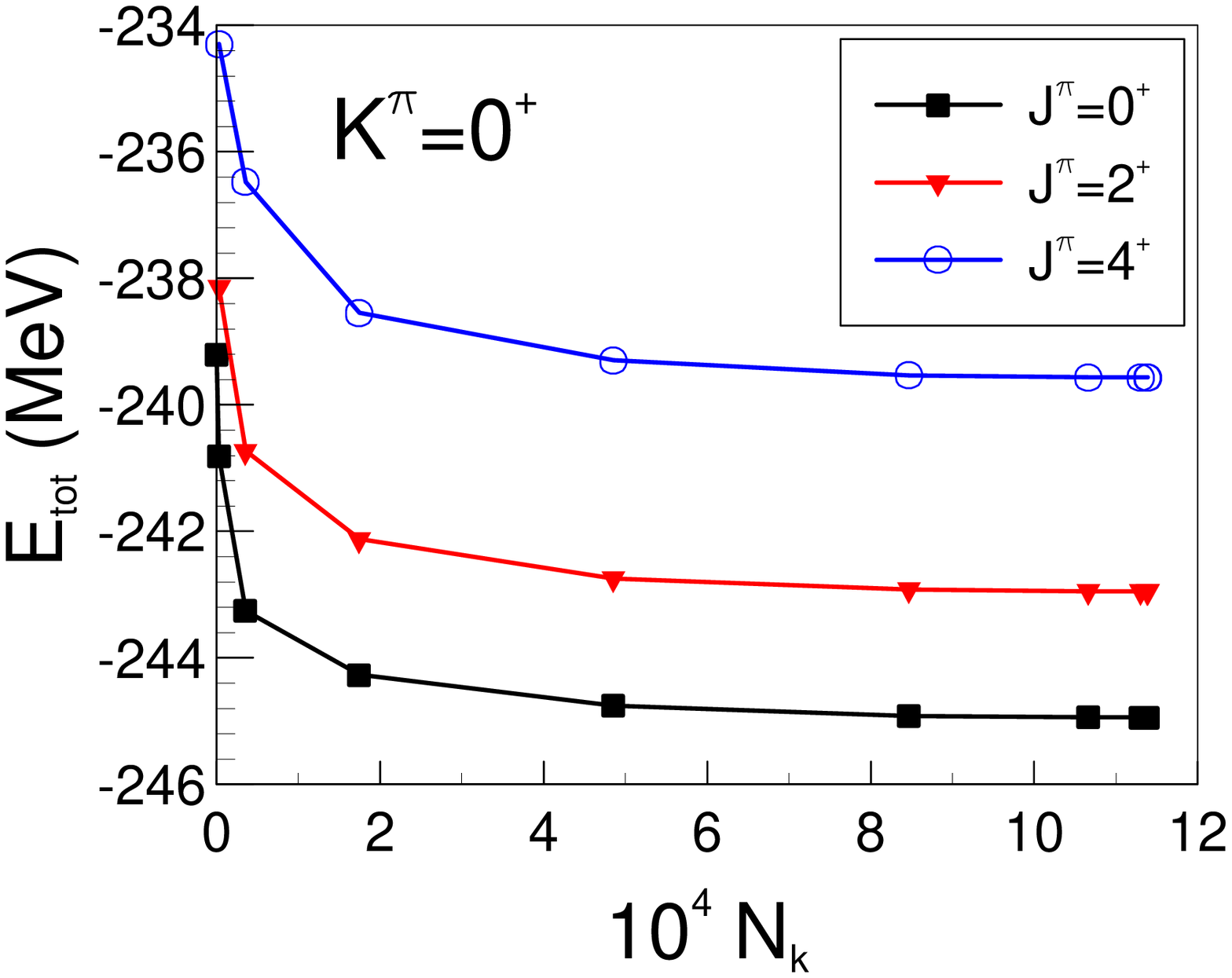}
\hspace{0.0cm}\includegraphics[height=6.5cm, angle=0]{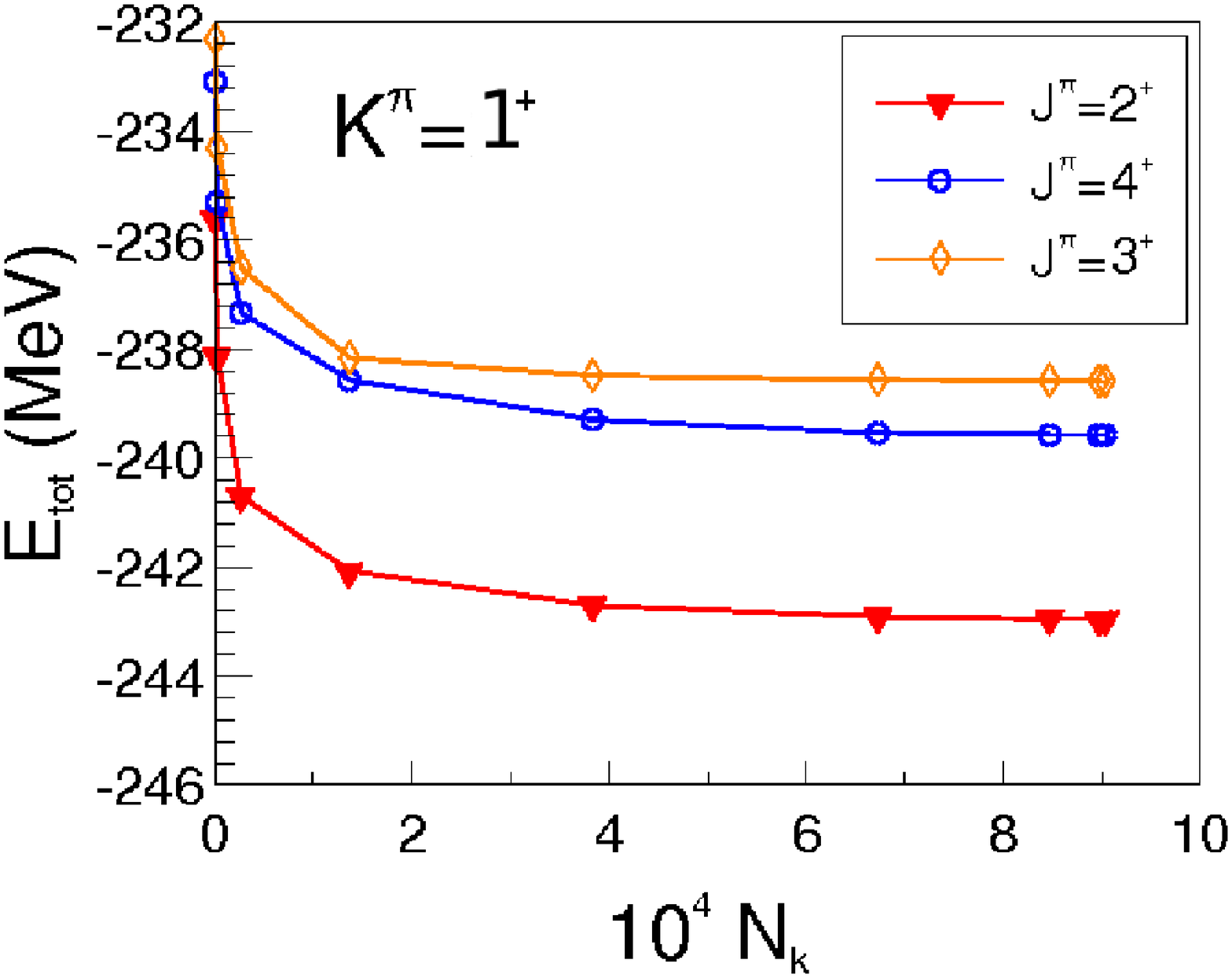}
\hspace{0.0cm}\includegraphics[height=6.5cm, angle=0]{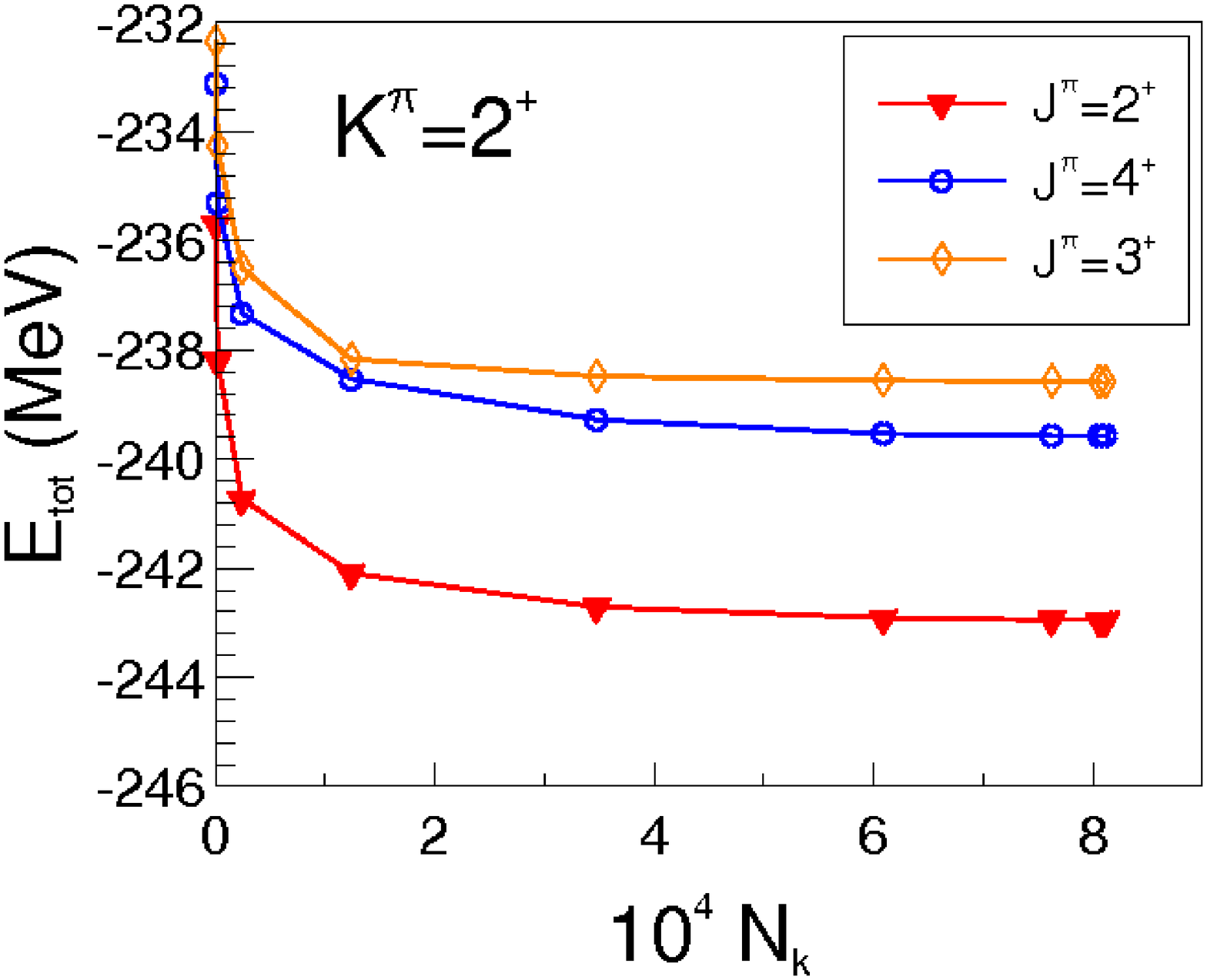}
\hspace{0.0cm}\includegraphics[height=6.5cm, angle=0]{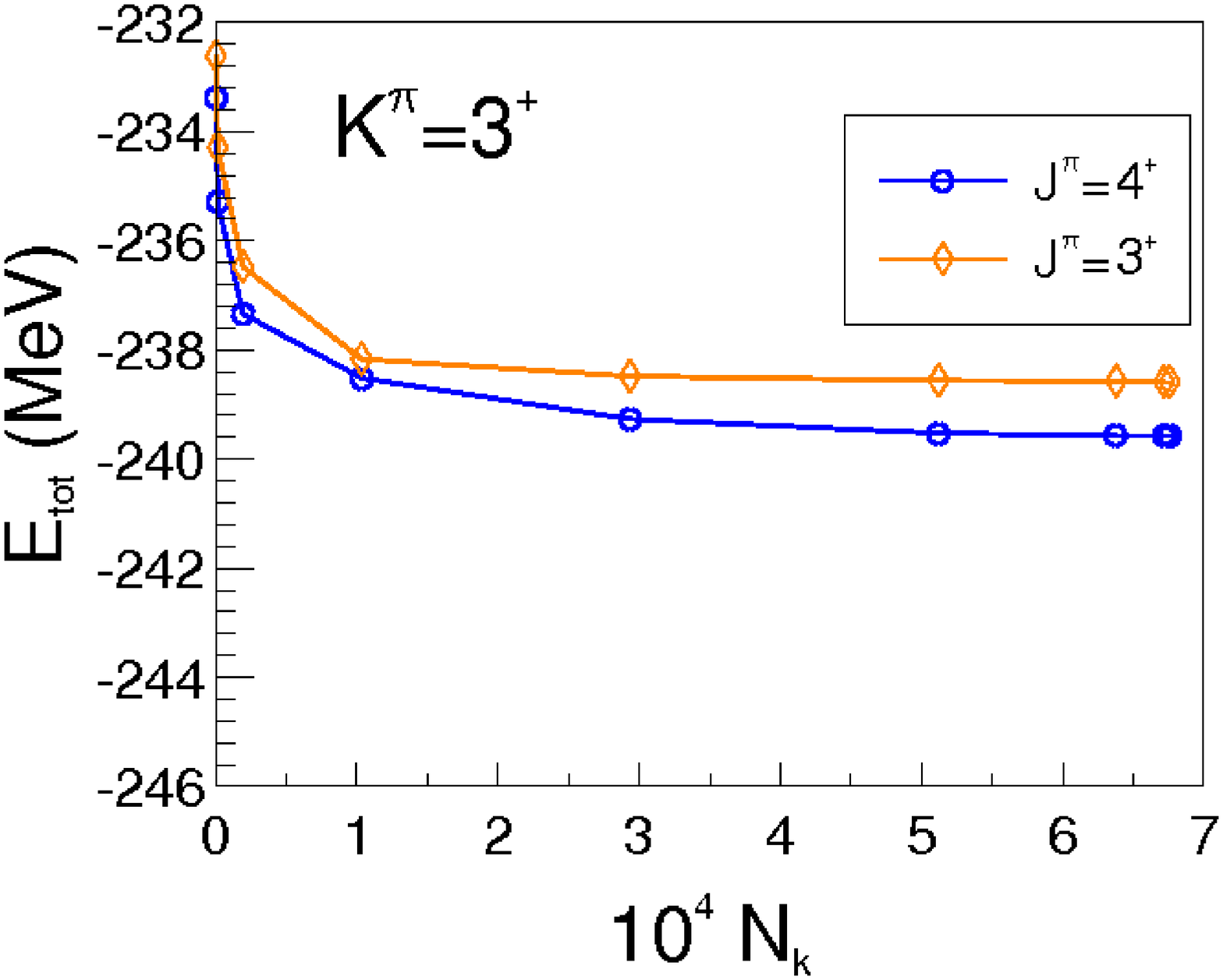}
\end{center}
\caption{(Color online) Convergence of ground state and a few excited state energies in $^{28}$Si
as a function of the number of configurations ordered by their centroid energies.}
\label{fig20}
\end{figure*}

In the case of the $K=0$ projection, one observes two unusual bumps at energies $E_{\alpha} -E_{k} \sim$ 40 MeV and
60 MeV. Actually, the strength function (\ref{st1}) combines two types of information: the density
of configurations and the mixing coefficients whose values are determined by the nucleon-nucleon interaction.
In order to disentangle these effects, we
calculate the density of configurations
$\rho(E)= \sum_{k, \alpha} \delta(E-{\cal E}_{\alpha})$ and the locally
normalized strength function $\langle F_{k}(E) \rangle/\rho(E)$ where
$\langle F_{k}(E) \rangle = \sum_{k} F_{k}(E) / N$. These quantities are
shown in Fig. \ref{fig7c} (the density of configurations, upper panel (a), and the locally
normalized strength
function in logarithmic scale, lower panel (b)). The configuration density has a regular shape expected
for a finite Hilbert space whereas the bumps of the strength function in Fig. \ref{fig7a} appear magnified on the normalized strength
function plot (lower panel).
One also notices that
the coupling of matrix elements is fully exponential (linear on logarithmic
scale), if one forgets about the bumps. As the density of configuration is
quasi-linear in logarithmic scale at large energy, one understands the
exponential convergence behavior of the strength function observed on Fig.
\ref{fig7a}.
The two bumps come from the specific interaction rather than from
the statistics.
The matrix elements that are responsible for the bumps correspond
to the same proton-neutron ME discussed in section \ref{lls}
for $^{30}$Si, which appear to be too
large at the level of the approximate
resolution of the $mp-mh$ configuration mixing method presented in this study.
Reducing by hand the values of these proton-neutron ME makes the bumps disappear.
Hence, the study of the chaotic behaviour of highly excited
Slater determinants can be considered as an additional way of highlighting not
well-calibrated coupling ME. In other words, statistical properties may serve as
an additional criterion in the validation process of
phenomenological effective interactions.
This result is consistent with and confirms our previous discussion
of Section \ref{lls}. This analysis is similar in spirit to
the statistical search of interaction matrix elements responsible for the
equilibrium prolate deformation \cite{horoi10}.

The exponential convergence observed in Fig. \ref{fig7c} is an interesting feature that might be evaluated analytically in the particular case of the finite-range
Gogny interaction. For example, taking the single-particle states as plane waves, the two-body matrix elements of the Brink-Boecker part of the Gogny force read:
\begin{equation}
\begin{array}{c}
\dspt \langle \phi_1 \phi_2 \vert e^{-(\vec{r}_1 -\vec{r}_2)^{2}/\mu^2} \vert \phi_3 \phi_4 \rangle \\
\dspt =\int \frac{d^{3}r_1\, d^{3}r_2}{(2 \pi)^6}\,
e^{-i \vec{k}_1 \cdot\vec{r}_1 } e^{-i \vec{k}_2 \cdot\vec{r_2} } e^{-(\vec{r}_1 -\vec{r}_2)^{2}/\mu^2}
e^{i \vec{k}_3 \cdot\vec{r}_1 } e^{i \vec{k}_4 \cdot\vec{r}_2 } \\
= \int \frac{d^{3}r_1\, d^{3}r_2}{(2 \pi)^6}\,
e^{-i (\vec{k}_1-\vec{k}_2)\cdot(\vec{r}_1-\vec{r}_2)}  e^{-(\vec{r}_1 -\vec{r}_2)^{2}/\mu^2}.
\end{array}
\label{w1}
\end{equation}
where we have used the conservation law $\vec{k}_1 + \vec{k}_2= \vec{k}_3+ \vec{k}_4$.

Eq. (\ref{w1}) reduces to the Fourier transform of a Gaussian,
\begin{equation}
\dspt \langle \phi_1 \phi_2 \vert e^{-(\vec{r}_1 -\vec{r}_2)^{2}/\mu^2} \vert \phi_3 \phi_4 \rangle = \frac{\mu}{\sqrt{2}}\,e^{-\frac{m \mu^2}{2} E},
\label{w3}
\end{equation}
with $E= k^2 / 2m$ and $\vec{k} = \vec{k}_1-\vec{k}_2 $ . Thus, the matrix element behaves, in the case of the Gogny interaction, as a decreasing
exponential with respect to excitation energy. The value of the two ranges introduced in the Gogny interaction may serve as a guide to decide
an upper limit where the exponential convergence regime settles. However, this behavior exists for any physically reasonable interaction as revealed in shell model calculations \cite{ec1,ec2,ec3,ec4,ec5}.

Below we give three examples of exponential convergence behavior with increasing excitation energy
of configurations, namely for the total energy, for the components of the correlated wave functions and for the occupation probabilities.
Fig. \ref{fig20} displays the evolution of the total energies of the ground and
excited states in $^{28}$Si, according to the number of configurations in the correlated wave function, ordered by increasing centroid energies.
The four plots correspond to the different values of the total momentum projection $K$.
In all cases, the total energy changes rapidly when only a few Slater determinants with lowest energy are included. Then, for a larger number of
configurations a smooth regime settles, a behavior independent of the value of $K$.

\begin{figure*}[htb!]
\begin{center}
\includegraphics[height=6.5cm, angle=0]{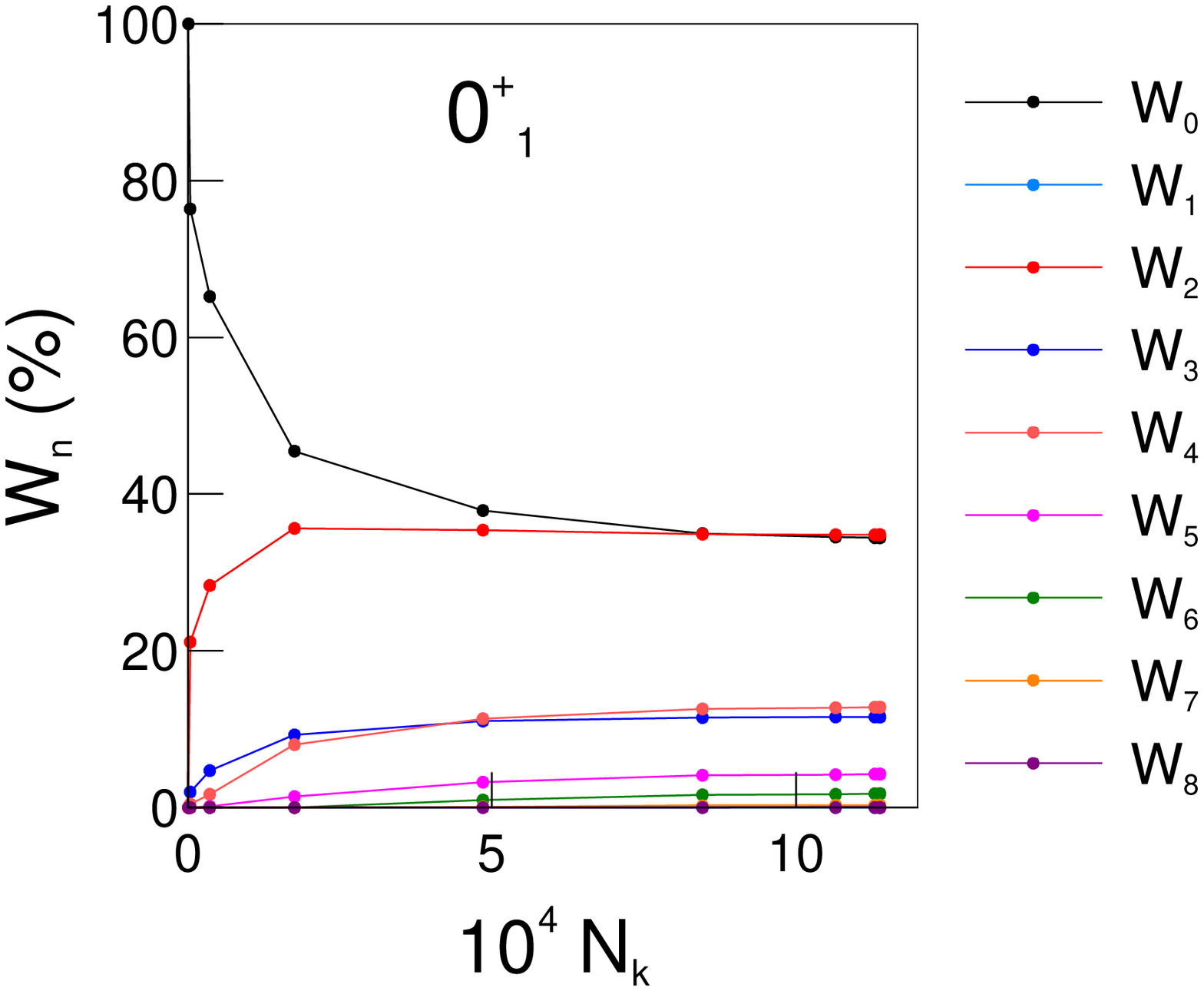}
\includegraphics[height=6.5cm, angle=0]{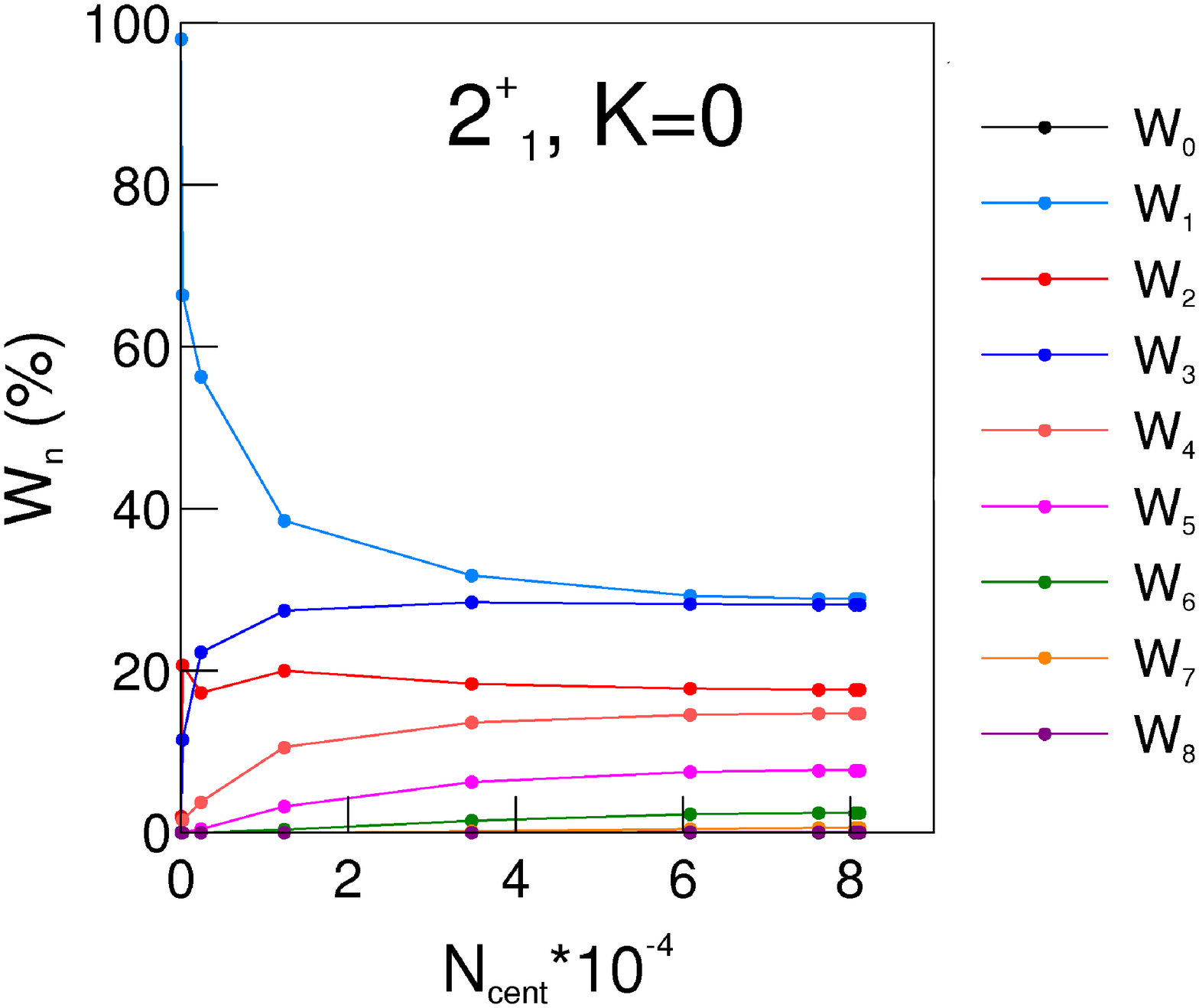}
\includegraphics[height=6.5cm, angle=0]{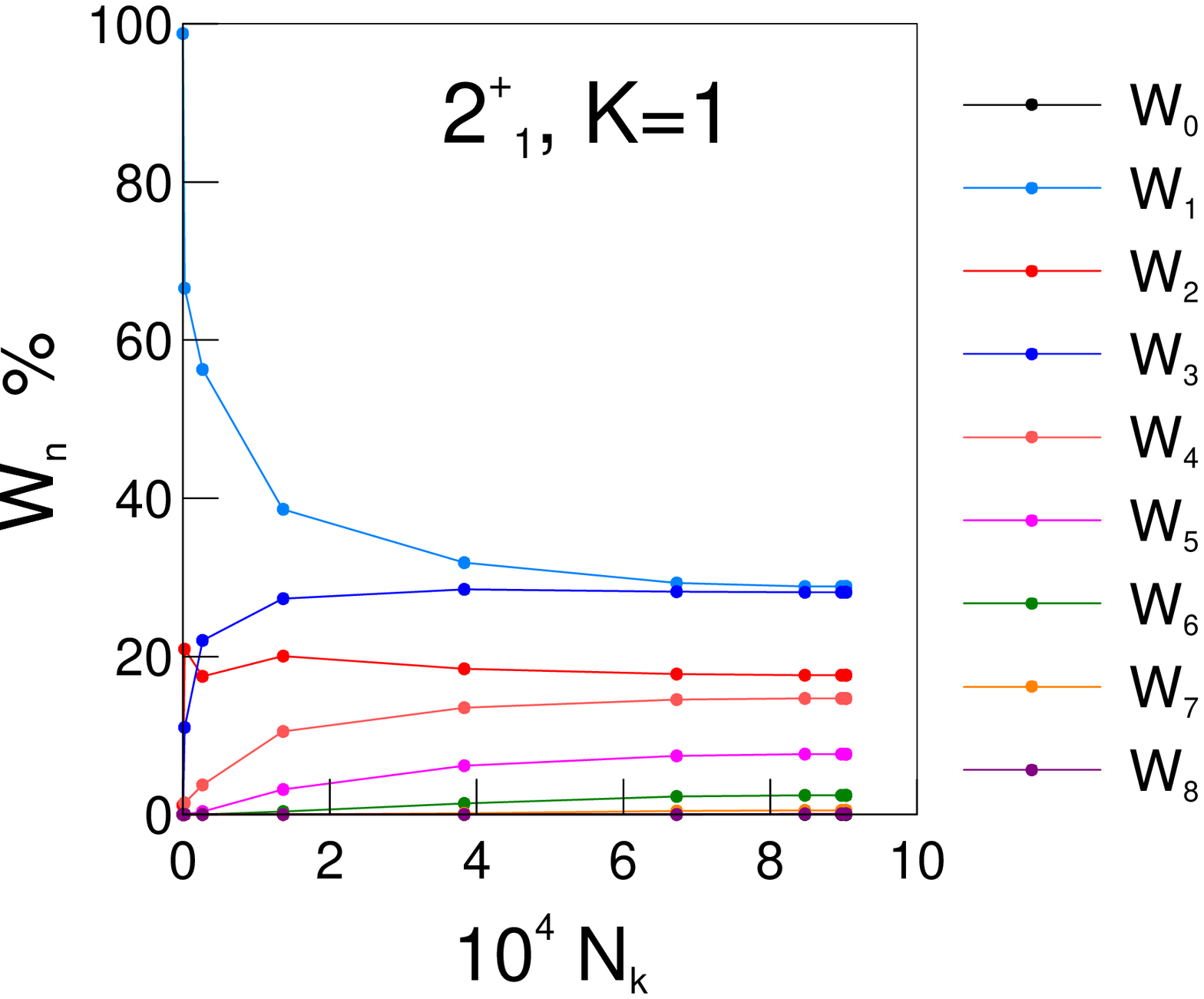}
\includegraphics[height=6.5cm, angle=0]{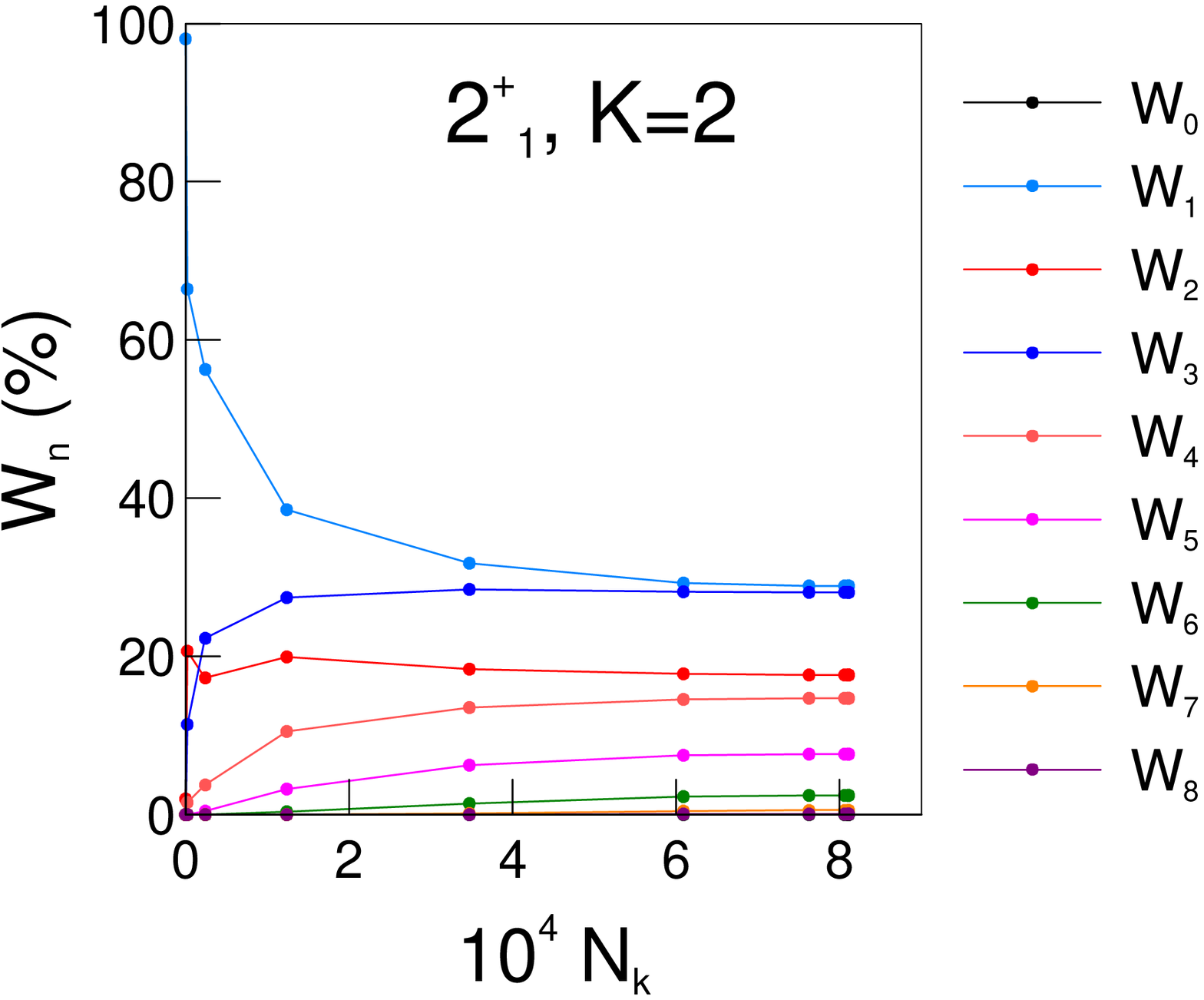}
\end{center}
\caption{(Color online) The $W_{i}$ components (Eq. (\ref{eqqq1})) of $0^+_1$ and $2^+_1$ wave functions in $^{28}$Si.}
\label{figSiwf}
\end{figure*}

\begin{figure}[htb!]
\begin{center}
\includegraphics[height=7.0cm, angle=0]{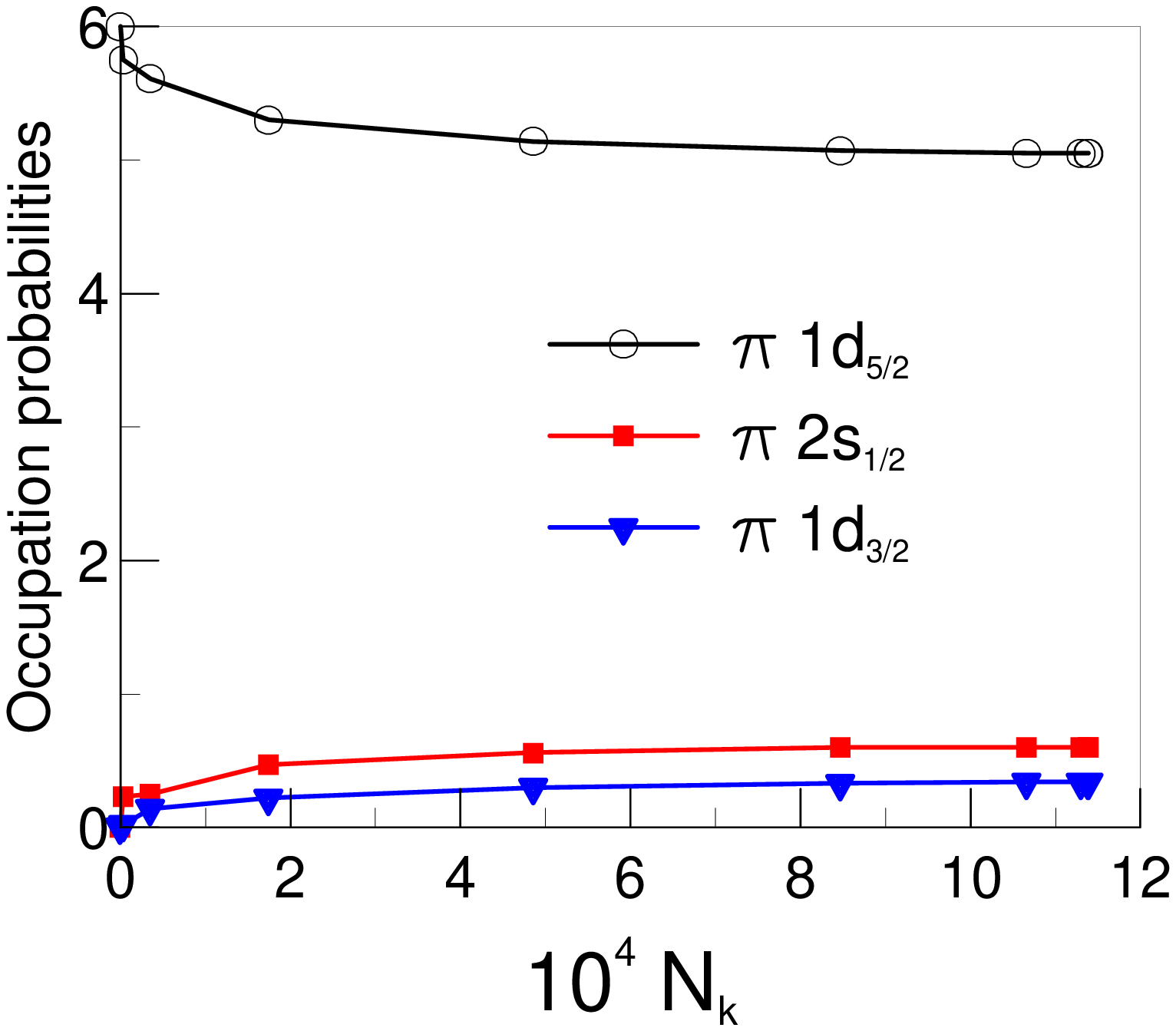}
\end{center}
\caption{(Color online) Evolution of the occupation probabilities of proton $d_{5/2}$, $s_{1/2}$ and $d_{3/2}$ orbitals as function of $N_{k}$.}
\label{figocc}
\end{figure}

In Fig. \ref{figSiwf}, the evolution of the global components, $W_{n}$, Eq. (\ref{eqqq1}), of the wave functions
of the $0^+_1$ and $2^+_1$ states in $^{28}$Si is displayed as a function of the number $N_{{\rm cent}}$ of centroids ordered by centroid energy.
For $n\geq$9, the $W_{n}$ are smaller than 10$^{-9}$ and they are not shown.
The evolution of the occupation probabilities for $d_{5/2}$, $s_{1/2}$ and $d_{3/2}$ orbitals according
to $N_{{\rm cent}}$ is displayed on Fig. \ref{figocc}.
As can be seen from the two figures, both the $W_{n}$ and the occupation probabilities display the exponential convergence.
With all exponents being close, one can loosely interpret this behavior as a signature of thermalization
in a self-bound mesoscopic system.

\section{Conclusion and perspectives}\label{section4}

In this work, we have investigated the application of multiconfiguration methods for the description
of low-energy nuclear spectroscopy. A few even-even silicon isotopes have been studied,
using the D1S Gogny interaction. At this stage, only the configuration mixing part of the method has been put into place;
the renormalization of single-particle orbitals due to correlations has not been discussed in this paper.
The results for the positive parity states in low-lying spectroscopy of $^{26-32}$Si have been found
in rather good agreement with experiment, taking into account the fact that the D1S Gogny interaction has not been
{\it{a priori}} fitted to be employed in such a kind of approach. In particular, from the study of $^{30}$Si,
it has been found that, at the level of approximate resolution of the $mp-mh$ configuration mixing method (no renormalisation of orbitals) a few residual
proton-neutron matrix elements of pairing type, that are not constrained in the fitting of the Gogny interaction, might disturb the reproduction of
excitation energies. In relation to this, the importance and magnitude of the proton-neutron residual interaction has been discussed.

The question of the pertinent configurations that have to be introduced in the mixing has also been addressed.
Statistical generic behavior of highly excited configurations has been put forward. In particular, the exponential
convergence already revealed in shell model studies has been confirmed in our approach.
This is an encouraging feature that may help to handle the very large number
of configurations that appear in multiconfiguration methods, in particular in nuclear physics where two kinds of particles exist.

The present study proposes interesting and challenging issues at different levels.
The renormalization of orbitals under the influence of correlations is an important question that will be analyzed in further studies.
In atomic physics, this renormalization has been proved to play a key role in strongly correlated systems. Even though the associated orbital equation,
see Eq. (\ref{eqad1}), looks simple, its exact solution is far from being trivial. At present, in the most
advanced applications to atomic physics, it is solved approximately as the correlation term that
depends on the two-body density matrix can be very complicated.

A second issue concerns the improvement of the Gogny interaction in order to be able to use it not only in HFB, RPA and GCM-type methods but also in
multiconfiguration approaches. Work is in progress in this direction.

The third issue deals with the generic behavior of highly excited configurations. The exponential convergence
and corresponding extrapolations can be of considerable help for controlling in a safe way possible new truncation schemes introducing
explicitly only pertinent configurations. To this aim, a formalism of the Feshbach type projection might be quite
useful. It is worth mentioning that the analysis of statistical regularities allowed us to identify specific
matrix elements of the Hamiltonian responsible for spectroscopic inadequacy (discussion on $^{30}$Si).

For the specific goal of nuclear spectroscopy, the investigation of  transition
probabilities and negative parity states in silicon isotopes
would be of great interest and may provide essential information on the residual interaction and properties
of single-particle orbitals, including nuclei far from stability.

\begin{acknowledgments}
N.P. would like to thank the National Superconducting Cyclotron Laboratory at Michigan State University for hospitality and B.A. Brown for providing the shell model matrix elements and results that were of great help for the analysis
presented in this work. N.P. is also grateful to M. Girod and J. Libert for providing with 5DCH results and A. Zuker and M. Dufour for interesting discussions.
V.Z. acknowledges the support from the NSF grants PHY-0758099 and PHY-1068217. Part of the calculations were carried out on CCRT
supercomputers of CEA-DAM Ile de France.
\end{acknowledgments}

\appendix

\section{ "Core + valence space" formulation} \label{a1}

Eqs. (\ref{eq2}) and (\ref{eq4}) are quite general. As mentioned in the Introduction, the Hilbert space
has to be truncated in any realistic calculation. Several truncation schemes, supported by physics arguments,
can be utilized, for example a limitation on the excitation order of $mp-mh$ configurations, a limitation in the number of
single-particle states used for the configuration mixing, etc. This appendix deals with a
truncation that corresponds to the description in terms of ``core + valence space". In this approach,
the system, that comprises $N^{\tau}$ nucleons of each isospin, is separated into: \\
- an even-even core where the $N_c^{\tau}$ lower individual states are fully
occupied for each isospin; \\
- a set of active orbitals containing $N^{\tau}-N_c^{\tau}$ particles for each isospin; \\
- a set of unoccupied higher-energy individual states for each isospin.

Under this prescription, the proton and neutron Slater determinants, Eq. (\ref{eq3}), are defined as
\begin{equation}
\dspt \vert \Phi_{\alpha_{\tau}} \rangle = \prod_{j=N_{c}^{\tau}+1}^{N^{\tau}} a^{+}_{j}
\prod_{i=1}^{N_{c}^{\tau}} a^{+}_{i} \vert 0 \rangle = \prod_{j=N_{c}^{\tau}+1}^{N^{\tau}}
a^{+}_{j} \vert \Phi_{c_{\tau}}\rangle,
\label{eqq1}
\end{equation}
where $\vert \Phi_{c_{\tau}}\rangle$ is the part of the wave function that describes the core.

In addition, the Hamiltonian $\hat{H} (\rho)$, Eq. (\ref{eq1}), can be written by separating isospin contributions as
\begin{equation}
\dspt \hat{H} (\rho) = \sum_{\tau = \pi, \nu} \hat{H}^{\tau} + \hat{V}^{\pi \nu} (\rho).
\label{eqq2}
\end{equation}
Then, from Eq. (\ref{eq5}),
\begin{equation}
\dspt \hat{{\cal{H}}} (\rho) = \sum_{\tau = \pi, \nu} [ \hat{H}^{\tau} + \sum_{mn \tau}
{\cal{R}}^{\tau}_{mn} a^{+}_{\tau m} a_{\tau n} ]+ \hat{V}^{\pi \nu} (\rho).
\label{eqq3b}
\end{equation}
where the generalized rearrangement coefficients ${\cal{R}}_{i j}^{\tau}$ are given by
\begin{equation}
\dspt {\cal{R}}_{m n}^{\tau} = \int \phi^*_{\tau m} (\vec{r}, \sigma) \phi_{\tau n} (\vec{r},
\sigma) B(\vec{r}) d^3 {\vec{r}},
\label{e655}
\end{equation}
with $\phi_{\tau i}$ the single-particle wave functions. The difficulties in the calculation of
${\cal{R}}_{m n}^{\tau}$ come from
the evaluation of the rearrangement field $B(\vec{r})$ from the correlated wave function (\ref{eq2}):
\begin{equation}
\begin{array}{c}
\dspt B(\vec{r}) = \sum_{\alpha \alpha'} A^*_{\alpha'_{\pi} \alpha'_{\nu}}
A_{\alpha_{\pi} \alpha_{\nu}}
\dspt \sum_{ijkl} \langle i_{\alpha'_{\pi}}
j_{\alpha'_{\nu}} \vert
\frac{\partial V (\rho)}{\partial \rho(\vec{r})} \vert \widetilde{k_{\alpha_{\pi}}
l_{\alpha_{\nu}}} \rangle \\
\times  \langle \Phi_{\alpha'_{\pi}} \vert a^+_{i_{\alpha'_{\pi}}} a_{k_{\alpha_{\pi}}}
\vert \Phi_{\alpha_{\pi}} \rangle
\langle \Phi_{\alpha'_{\nu}} \vert a^+_{j_{\alpha'_{\nu}}} a_{l_{\alpha_{\nu}}}
\vert \Phi_{\alpha_{\nu}} \rangle
\end{array}
\label{e889}
\end{equation}
The expressions (\ref{eqq2}), (\ref{eqq3b}) and (\ref{e889}) are specific to the D1S Gogny interaction as
only the proton-neutron terms are generated by the density-dependent part of the interaction and the density-dependence is a contact interaction.
In the limit of no configuration mixing, one recovers the standard HF expression for
the rearrangement term.

From Eq. (\ref{eqq3b}), one derives explicit expressions for $ {\cal{H}}_{\alpha_{\pi} \alpha_{\nu},\alpha'_{\pi}
\alpha'_{\nu} }$ in terms of the ``core + valence space" formulation. In the following evaluation of
$ {\cal{H}}_{\alpha_{\pi} \alpha_{\nu},\alpha'_{\pi} \alpha'_{\nu} }$, only the terms carrying the core
contributions are given.

\subsection{Proton and neutron diagonal contributions}

The proton and neutron diagonal contributions are obtained for $\vert \Phi_{\alpha_{\pi}} \rangle \equiv \vert
\Phi_{\alpha'_{\pi}} \rangle$ and $\vert \Phi_{\alpha_{\nu}} \rangle \equiv \vert \Phi_{\alpha'_{\nu}} \rangle$ as
\begin{equation}
\dspt  {\cal{H}}^{\tau}_{\alpha_{\pi} \alpha_{\nu},\alpha_{\pi} \alpha_{\nu} } = \langle \phi_{\alpha_{\tau}}
\vert \hat{H}^{\tau} \vert \phi_{\alpha_{\tau}} \rangle +
\sum_{m n} {\cal{R}}_{mn}^{\tau} \langle \phi_{\alpha_{\tau}}
\vert a^+_m a_n \vert \phi_{\alpha_{\tau}} \rangle.
\label{e80}
\end{equation}
which yields
\begin{equation}
\begin{array}{c}
\dspt  {\cal{H}}^{\tau}_{\alpha_{\pi} \alpha_{\nu},\alpha_{\pi} \alpha_{\nu} } =
\sum_{i_{\alpha_{\tau}}=1}^{N^{\tau}} (\langle i_{\alpha_{\tau}}
\vert K \vert i_{\alpha_{\tau}} \rangle +
{\cal{R}}_{i_{\alpha_{\tau}} i_{\alpha_{\tau}}}^{\tau}) \\
\dspt + \frac{1}{2}
\sum_{i, j = 1}^{N^{\tau}}
\langle i_{\alpha_{\tau}} j_{\alpha_{\tau}} \vert V(\rho) \vert
\widetilde{i_{\alpha_{\tau}} j_{\alpha_{\tau}}} \rangle.
\end{array}
\label{e81}
\end{equation}
In Eq. (\ref{e81}), $i_{\alpha_{\tau}}$ stands for the occupied single-particle state in
the $\vert \Phi_{\alpha_{\tau}} \rangle$ Slater determinant. The ``core + valence space" separationthen  leads to:
\begin{equation}
\begin{array}{c}
\dspt  {\cal{H}}^{\tau}_{\alpha_{\pi} \alpha_{\nu},\alpha_{\pi} \alpha_{\nu} } =
\sum_{i_{\alpha_{\tau}}=1}^{N_{c}^{\tau}} (\langle i_{\alpha_{\tau}} \vert K \vert i_{\alpha_{\tau}} \rangle +
{\cal{R}}_{i_{\alpha_{\tau}} i_{\alpha_{\tau}}}^{\tau})  \\
\dspt + \sum_{i_{\alpha_{\tau}}=N_{c}^{\tau}+1}^{N^{\tau}} (\langle i_{\alpha_{\tau}} \vert K \vert i_{\alpha_{\tau}} \rangle +
{\cal{R}}_{i_{\alpha_{\tau}} i_{\alpha_{\tau}}}^{\tau}) \\
\dspt + \frac{1}{2} \sum_{i_{\alpha_{\tau}}=1}^{N_{c}^{\tau}} \sum_{j_{\alpha_{\tau}}=1}^{N_{c}^{\tau}}
\langle i_{\alpha_{\tau}} j_{\alpha_{\tau}} \vert V \vert \widetilde{i_{\alpha_{\tau}}
j_{\alpha_{\tau}}}  \rangle \\
\dspt + \frac{1}{2} \sum_{i_{\alpha_{\tau}}=N_{c}^{\tau}+1}^{N^{\tau}}
\sum_{j_{\alpha_{\tau}}=N_{c}^{\tau}+1}^{N^{\tau}}
\langle i_{\alpha_{\tau}} j_{\alpha_{\tau}} \vert V \vert \widetilde{i_{\alpha_{\tau}}
j_{\alpha_{\tau}}}  \rangle \\
\dspt + \sum_{i_{\alpha_{\tau}}=1}^{N_{c}^{\tau}}
\sum_{j_{\alpha_{\tau}}=N_{c}^{\tau}+1}^{N^{\tau}}
\langle i_{\alpha_{\tau}} j_{\alpha_{\tau}} \vert V \vert
\widetilde{i_{\alpha_{\tau}} j_{\alpha_{\tau}}}  \rangle.
\end{array}
\label{eqq4}
\end{equation}
In Eq. (\ref{eqq4}), the first and third terms are pure core contributions. The second and forth terms
are the contributions from the valence space. The mixed fifth term includes single-particle orbitals
of both the core and the valence space.

\subsection{Proton-neutron diagonal contribution}

The proton-neutron diagonal contribution is obtained for $\vert \Phi_{\alpha_{\pi}}
\rangle \equiv \vert \Phi_{\alpha'_{\pi}} \rangle$ and $\vert \Phi_{\alpha_{\nu}} \rangle \equiv \vert
\Phi_{\alpha'_{\nu}} \rangle$, as
\begin{equation}
\dspt {\cal{H}}_{\alpha_{\pi} \alpha_{\nu},\alpha_{\pi} \alpha_{\nu}}^{\pi \nu} =
\langle \Phi_{\alpha_{\pi}} \Phi_{\alpha_{\nu}} \vert \hat{V}^{\pi \nu} \vert
\Phi_{\alpha_{\pi}} \Phi_{\alpha_{\nu}} \rangle.
\label{e84}
\end{equation}
Expansion of Eq. (\ref{e84}) yields
\begin{equation}
\dspt {\cal{H}}_{\alpha_{\pi} \alpha_{\nu},\alpha_{\pi} \alpha_{\nu}}^{\pi \nu} =
\sum_{i=1}^{N^{\pi}} \sum_{j=1}^{N^{\nu}}
\langle i_{\alpha_{\pi}} j_{\alpha_{\nu}} \vert V \vert \widetilde{i_{\alpha_{\pi}}
j_{\alpha_{\nu}}} \rangle.
\label{e85}
\end{equation}
Making the "core + valence space" separation leads to
\begin{equation}
\begin{array}{c}
\dspt {\cal{H}}_{\alpha_{\pi} \alpha_{\nu},\alpha_{\pi} \alpha_{\nu}}^{\pi \nu} =
\sum_{i_{\alpha_{\pi}}=1}^{N^{\pi}_c} \sum_{j_{\alpha_{\nu}}=1}^{N^{\nu}_c}
\langle i_{\alpha_{\pi}} j_{\alpha_{\nu}} \vert V \vert
\widetilde{i_{\alpha_{\pi}} j_{\alpha_{\nu}}} \rangle \\
\dspt + \sum_{i_{\alpha_{\pi}}=N^{\pi}_c+1}^{N^{\pi}} \sum_{j_{\alpha_{\nu}}=N^{\nu}_c+1}^{N^{\nu}}
\langle i_{\alpha_{\pi}} j_{\alpha_{\nu}} \vert V \vert \widetilde{i_{\alpha_{\pi}}
j_{\alpha_{\nu}}} \rangle \\
\dspt + \sum_{i_{\alpha_{\pi}}=1}^{N^{\pi}_c} \sum_{j_{\alpha_{\nu}}=N^{\nu}_c+1}^{N^{\nu}}
\langle i_{\alpha_{\pi}} j_{\alpha_{\nu}} \vert V \vert \widetilde{i_{\alpha_{\pi}}
j_{\alpha_{\nu}}} \rangle \\
\dspt + \sum_{i_{\alpha_{\pi}}=N^{\pi}_c+1}^{N^{\pi}} \sum_{j_{\alpha_{\nu}}=1}^{N^{\nu}_c}
\langle i_{\alpha_{\pi}} j_{\alpha_{\nu}} \vert V \vert \widetilde{i_{\alpha_{\pi}}
j_{\alpha_{\nu}}} \rangle.
\end{array}
\label{e86}
\end{equation}
In Eq. (\ref{e86}), the first term comes from the core and the second one from the valence space. The last two terms
express the coupling between the core and the valence particles.

\subsection{Proton and neutron non-diagonal one-body contributions}

The proton or neutron one-body contributions arise when
$\vert \Phi_{\alpha_{\pi}} \rangle$ and $\vert \Phi_{\alpha'_{\pi}} \rangle$ differ in one particle state and
$\vert \Phi_{\alpha_{\nu}} \rangle \equiv \vert \Phi_{\alpha'_{\nu}} \rangle$ or when
$\vert \Phi_{\alpha_{\nu}} \rangle$ and $\vert \Phi_{\alpha'_{\nu}} \rangle$ differ from one particle state and
$\vert \Phi_{\alpha_{\pi}} \rangle \equiv \vert \Phi_{\alpha'_{\pi}} \rangle$, respectively.
Then, the contributions to ${\cal{H}}$ are
\begin{equation}
\begin{array}{c}
\dspt {\cal{H}}_{\alpha_{\pi} \alpha_{\nu},\alpha'_{\pi} \alpha_{\nu}}^{1} =
\sum_{i_{\alpha_{\pi}}=1}^{N_{c}^{\pi}} \langle i_{\alpha_{\pi}}
j_{\alpha'_{\pi}} \vert V \vert \widetilde{ i_{\alpha_{\pi}}
l_{\alpha'_{\pi}} } \rangle \\
\dspt + \sum_{i_{\alpha_{\pi}}=N_{c}^{\pi}+1}^{N^{\pi}} \langle
i_{\alpha_{\pi}} j_{\alpha'_{\pi}} \vert V \vert \widetilde{ i_{\alpha_{\pi}}
l_{\alpha'_{\pi}} } \rangle
\end{array}
\label{eqq5}
\end{equation}
and
\begin{equation}
\begin{array}{c}
\dspt {\cal{H}}_{\alpha_{\pi} \alpha_{\nu},\alpha_{\pi} \alpha'_{\nu}}^{1} =
\sum_{i_{\alpha_{\nu}}=1}^{N_{c}^{\nu}} \langle i_{\alpha_{\nu}}
j_{\alpha'_{\nu}} \vert V(\rho) \vert \widetilde{ i_{\alpha_{\nu}}
l_{\alpha'_{\nu}} } \rangle \\
\dspt + \sum_{i_{\alpha_{\nu}}=N_{c}^{\nu}+1}^{N^{\nu}} \langle
i_{\alpha_{\nu}} j_{\alpha'_{\nu}} \vert V(\rho) \vert \widetilde{ i_{\alpha_{\nu}}
l_{\alpha'_{\nu}} }\rangle.
\end{array}
\label{eqq5b}
\end{equation}
In Eqs. (\ref{eqq5})-(\ref{eqq5b}), the indices $j$ and $l$ refer to the single-particle state different
between the configurations $\alpha$ and $\alpha'$. The first term in Eqs. (\ref{eqq5})-(\ref{eqq5b}) corresponds
to the core contribution and the second one to the contribution of the valence space.

\subsection{Proton-neutron non-diagonal one-body contribution}

As for the proton and neutron one-body contributions, the proton-neutron one-body contribution arises
when $\vert \Phi_{\alpha_{\pi}} \rangle$ and $\vert \Phi_{\alpha'_{\pi}} \rangle$
differ by one particle state and $\vert \Phi_{\alpha_{\nu}} \rangle \equiv \vert \Phi_{\alpha'_{\nu}} \rangle$ or when
$\vert \Phi_{\alpha_{\nu}} \rangle$ and $\vert \Phi_{\alpha'_{\nu}} \rangle$ differ by one particle state and
$\vert \Phi_{\alpha_{\pi}} \rangle \equiv \vert \Phi_{\alpha'_{\pi}} \rangle$.
Similarly to Eqs. (\ref{eqq5})-(\ref{eqq5b}), the proton-neutron non-diagonal one-body contribution is then given by
\begin{equation}
\begin{array}{c}
\dspt {\cal{H}}_{\alpha_{\tau} \alpha_{\tau'},\alpha_{\tau} \alpha'_{\tau'}}^{1~\tau'} =
\sum_{i_{\alpha_{\tau}}=1}^{N_{c}^{\tau}} \langle i_{\alpha_{\tau}} j_{\alpha_{\tau'}}
\vert V^{\pi \nu}(\rho) \vert \widetilde{ i_{\alpha_{\tau}}
l_{\alpha'_{\tau'}} }\rangle \\
\dspt + \sum_{i_{\alpha_{\tau}}=N_{c}^{\tau}+1}^{N^{\tau}} \langle i_{\alpha_{\tau}}
j_{\alpha_{\tau'}} \vert V^{\pi \nu}(\rho) \vert \widetilde{i_{\alpha_{\tau}}
l_{\alpha'_{\tau}} }\rangle.
\end{array}
\label{eqq6}
\end{equation}
In Eq. (\ref{eqq6}), the indices $j$ and $l$ refer to the single-particle states that differ
between the configurations $\alpha$ and $\alpha'$. Still, the first term describes the core contribution and
the second one the contribution of the valence space.

\end{document}